\documentclass[useAMS,usenatbib]{mn2e}
\usepackage{epsfig} 
\usepackage{graphicx}
\usepackage{color} 
\usepackage{amsmath} 
\def\eg{{\it e.g.,}}
\def\etal{{\em et al.}}
\def\ie{{\em i.e.,}}
\def\lsim{\lower.5ex\hbox{$\; \buildrel < \over \sim \;$}}
\def\gsim{\lower.5ex\hbox{$\; \buildrel > \over \sim \;$}}
\def \simeq{\lower.3ex\hbox{$\; \buildrel \sim \over - \;$}}

\def\apj{{ApJ}}%
\def\apss{{Ap\&SS}}
\def\aap{{A\&A}}%
\def\mnras{{MNRAS}}%
\def\na{{New A}}%
\def\pasj{{PASJ}}%
\title[Mass outflow from viscous discs]
{Estimation of the mass outflow rates from viscous accretion discs}
\author[Kumar, R. and Chattopadhyay, I.]{Rajiv Kumar$^{1}$\thanks{E-mail: rajiv.k@aries.res.in (RK); indra@aries.res.in (IC)},
Indranil Chattopadhyay$^1$\\
$^{1}$Aryabhatta Research Institute of Observational Sciences (ARIES),
Manora Peak, Nainital $-$ 263129, India\\
 }
\begin{document}
\date{Accepted -----. Received ------; in original form ----}

\pagerange{\pageref{firstpage}--\pageref{lastpage}} \pubyear{2012}

\maketitle

\label{firstpage}
\begin{abstract}
We study viscous accretion disc around black holes, and all possible accretion solutions, including
shocked as well as shock free accretion branches.
Shock driven bipolar outflows from a viscous accretion disc around a black hole has been investigated.
One can identify two critical viscosity parameters $\alpha_{cl}$ and $\alpha_{cu}$, within
which the stationary shocks may occur, for each set of boundary conditions.
Adiabatic shock has been found for upto viscosity parameter $\alpha=0.3$, while
in presence of dissipation and massloss we have found stationary shock upto $\alpha=0.15$.
The mass outflow rate may increase
or decrease with the
change in disc parameters, and is usually around few to 10 \% of the mass inflow rate.
We show that for the same outer boundary condition, the shock
front decreases to a smaller distance with the increase of $\alpha$.
We also show that the increase in dissipation reduces the thermal
driving in the post-shock disc, and hence the mass outflow rate decreases upto a few \%.
\end{abstract}
%% each reference.
%%%%%%%%%%%%%%%%%%%SECTION 1 %%%%%%%%%%%%%%%%%%
\section{Introduction}

One of the curious thing about the galactic as well as extra galactic black hole
candidates is that, they show moderate to strong jets. Since black holes are compact and
do not have hard surfaces,
outflows and jets can originate only from the accreting material. And observations have indeed
showed that the persistent jet or outflow activities are correlated with the hard spectral states
of the disc \ie when the power radiated maximizes in the hard non-thermal part of the spectrum
\citep{gfp03}. Moreover, there seems to be enough evidence that these jets
are produced within 100 Schwarzschild radius of the central object \citep{jbl99}.
So accretion disc models which are to be considered for persistent jet generation should allow the
formation of jet
close to the central object, and that the spectra of the disc should be in the hard state.

One of the most widely used accretion disc model --- the Keplerian disc \citep{ss73,nt73},
is very successful in explaining the multicoloured black body part of the spectrum. However,
the presence of Keplerian disc alone is inadequate in explaining the existence of the non-thermal
tail of the spectrum. Moreover, the advection term in the equation of motion is poorly handled in this model, 
 where the inner disc was chopped off adhoc,
and cannot automatically explain jet generation mechanism, neither can it explain the relative
small size of the jet base. Hence, one has to consider accretion models which have significant
advection. An elegant summary of all types of viscous accretion disc model has been provided
by \citet{ldc11}. For the sake of completeness, we present a very brief account of advective discs, in the following.

Accretion onto black hole is necessarily transonic since
radial velocity can only be small far away from the black hole, but matter must cross the horizon with 
 velocity equal to the speed of light
($c$). Hence accretion solutions around black holes should consider
the advection term self consistently. Accretion solution like Bondi flow or, radial flow
\citep{b52,cr09} although
satisfies the transonicity criteria, but is of very low luminosity. Therefore, there was a need to consider
rotating flow which
are transonic too, such that the infall timescale would be long enough to generate the photons observed
from microquasars and AGNs. The accretion model with
advection which got wide attention, was ADAF \citep{i77,ny94}. Initially ADAF was constructed around a
Newtonian gravitational potential, where the viscously dissipated energy is advected along with the mass,
and the momentum of the flow. The original ADAF solution was self-similar and wholly subsonic,
thus violating the
inner boundary condition around a black hole. This inadequacy was restored when global solutions of ADAF
in presence of strong gravity
showed that the flow actually becomes transonic at around few Schwarzschild radii ($r_g$), while
remains subsonic elsewhere. The self-similarity of such a solution may be maintained far away from the
sonic point \citep{cal97}. Moreover, the Bernoulli parameter is positive making the solutions unbound.
This prompted the introduction of self-similar outflows in the ADAF abbreviated as ADIOS, which would
carry away mass, momentum and energy \citep{bb99}. Although this is an interesting addition to the
generalization of ADAF type solutions, no physical mechanism was identified except the
positivity of Bernoulli parameter of the accretion flow, that would drive these outflows.
%Bondi type flows also show positive Bernoulli parameter, these accretion solution do not
%show outflows untill and unless a proper physical mechanism drives them \citep{ke86}.

Simultaneous to the research on ADAF class of solutions, a lot of progress was being made in the research of
general advective, rotating solutions. For rotating advective flow, \citet{lt80} showed
that with the increase in angular momentum of the inviscid flow, the number of physical critical points increases
from one to two, and later it was shown that a standing shock can exist
in between the two critical points \citep{f87,c89,fk07,c08,cc11}. 
%Rotating accreting matter in the advective regime \citep{lt80, f87, c89, ct95} do satisfy the transonicity criteria, and since the matter is rotating, it spends
%long enough time to generate the high energy photons. For rotating advective flow, \citet{lt80} showed
%that with the increase in angular momentum of the flow, the number of physical critical points increases
%from one to two, and later it was shown that a standing shock can exist
%in between the two critical points \citep{f87,c89,fk07,c08,cc11}.
Since the accretion shock is
centrifugal pressure mediated, there was apprehension about the stability and formation of such shocks,  
in presence of processes such as viscosity, which transports angular momentum outwards.
All doubts about the stability of such shocks in presence of viscosity was subsequently
removed \citep{lmc98,c96,cd04,cd07,dc08,ldc11}. Furthermore, \citet{ft04} conjectured about the
existence of multiple shocks and later, independent numerical simulations serendipitously found
the existence of transient multiple shocks in presence of Shakura Sunyaev type
viscosity \citep{letal08,ldc11}. Although, both general advective and ADAF solutions
start with the same set of equations, it was intriguing that there can be two mutually exclusive
class of solutions,
and without any knowledge under which condition these solutions may arise.
\citet{lgy99} later showed that the global ADAF solution is a subset of general advective solutions.
In other words, the models which concentrate only on the regime 
where the gravitational energy is converted mainly to the thermal and rotational energy, may either be 
cooling dominated (\eg Keplerian disc; Shakura \& Sunyaev, 1973; Novikov \& Thorne, 1973) or advection dominated
\citep{nkh97}.
Either way, these models remain mainly subsonic (except very close to the black 
hole) and hence do not show shock transition. However, if the entire
parameter space is searched, one can retrieve solutions which have 
enough kinetic energy to become transonic at distances of few $\times ~ 100r_g$.
A subset amongst these solutions admit shock transitions when shock conditions
are satisfied. This is the physical reason why
some disk models show shock transitions and others do not \citep{lgy99,dbl09}.
Whether a flow will follow an ADAF solution or some kind of hybrid solution
with or without shock will depend on the outer boundary condition and the physical processes dominant in the disc.
%which was also confirmed by \citet{dbl09}.

The shock model was later used to explain observations. \citet{ct95,cm06,mc08,mc10} showed that the post-shock region being
hotter, can produce the hard power-law tail by inverse-Comptonizing soft photons from pre-shock and post-shock parts of the accretion
disc. The soft state and hard states are automatically explained depending on the existence or non-existence of the shock.
Presence of such transonic advective flow has also been suggested by observations \citep{shms01,shs02}.
Infact the `hardness-intensity-diagram', or HID a hysteresis like behaviour seen in microquasars has been reproduced by
this
model \citep{mc10}. In other words, the post-shock region is the elusive Compton cloud.

Interestingly enough, the shock fronts are stable in a limited region of energy-angular momentum parameter space
and naturally give rise to time dependent solutions whenever the exact momentum balance across the
shock front is not achieved. This might be due to different rates of cooling \citep{msc96}, or different rates
of viscous transport \citep{lmc98,ldc11}. These oscillations were also confirmed by pure general relativistic
simulations \citep{akkn04,ny08,ny09}. \citet{msc96} suggested that if the post-shock region oscillates,
then the hard radiation produced by the post-shock region would oscillate as well, and hence explain the
QPO. Infact, the evolution of QPO frequencies during the outburst states of various microquasars 
like XTEJ1550-654, GRO 1655-40 etc were explained by this model \citep{cdp09} by assuming
inward drift of the oscillating shock due to increased viscosity of the flow.

Another interesting consequence of accretion shock is that, it automatically explains
the formation of outflows. The unbalanced pressure gradient force in the axial direction
drives matter in the form of outflows and may be considered as the precursor of jets
\citep{mlc94,c99,cd07,dc08}. These outflows can be accelerated by various 
accelerating mechanisms to relativistic terminal speeds \citep{cdc04,c05}.
If the shock conditions are properly considered,
then the mass outflow should reduce the pressure of the post-shock region and hence
the shock would move inward, which in turn, would modify the shock parameter space in presence of
massloss \citep{sc12}. Another model of non-fluid bipolar outflows have been
enthusiastically pursued by Becker and his collaborators which involves Fermi acceleration of
particles in isothermal shocks \citep{lb05,bdl08,dbl09}. 
The advantage of shock in accretion model is that the presence or absence of shock, is good enough to broadly
explain
many varied aspects of black holes candidates such as spectral states, QPOs, jets and outflows etc and the correlation
between these aspects \citep{cdp09}. Such correlations are also reported in observations
\citep{shms01,shs02,gfp03}.

There are
other jet generation models too. Apart from ADIOS, magnetically driven outflows were also proposed by many authors
\citep{bz77,bp82,p05,hbk07}. 
These magnetic bipolar outflows may be powered by the extraction of rotation energy of the black hole
in the form of Poynting flux.  However, recent observations finds weak or no correlation of black hole spin with
the jet formation around microquasars and are conjectured to be similar for AGNs as well \citep{fgr10}.
Bipolar outflows may even be powered by centrifugal or magnetic effect, and has been
shown that for low angular momentum even weak magnetic fields can produce equatorial inflow, bipolar outflow, 
polar funnel inflow and polar funnel outflow, and magnetic effect was identified as the
main driver of such outflows. Interesting as it may be, however, its connection with spectral
and radiative state of black hole candidates is not well explored. It is well known that
the spectra of the black hole candidates extends to high energy domain ($\sim$ few MeVs),
and one of the ways to generate such high energy non-thermal spectra is by shock acceleration of
electrons, and has been used to explain spectra from black hole candidates \citep{cm06,mc08}.
Not only hydrodynamic calculations, even magneto-hydrodynamic investigations
bears the possibility of shock in accretion \citep{nrks05,tgfrt06}.
Therefore we are looking for solutions of outflow generation which incorporates shocks,
although presently investigating the effect of viscosity in formation of such outflows.

%Numerical simulations show the existence of thermally driven
%bipolar outflows from the post-shock
%disc, where the shocks are far from isothermal approximation \citep{lmc98,ldc11}.
Various simulations with Pacz\'ynski-Wiita potential \citep{mlc94,msc96,mrc96,lmc98,letal08},
as well as GRMHD simulations \citep{nrks05} showed the presence of post-shock bipolar outflows,
which are far from isothermal approximation. Therefore, presently we relax the strict isothermality
condition,
and concentrate on conservation of fluxes across the shock front.
Theoretical frame work of thermally driven outflows has been done either
for inviscid discs \citep{c99,dcc01,sc12}, or for viscous discs where the viscous stress
was assumed to be proportional to the total pressure
\citep{cd07,dc08}. No theoretical investigations has thus far been made to study
thermally driven outflows from discs where the viscous stress is proportional to the shear,
a form of viscosity which is probably more realistic for black hole system \citep{bs05}.
%And address one of the thorny issues of determining the sonic point in such a case.
In this paper we solve viscous accretion disc equations for the contentious viscosity prescription,
to compute the solution topologies of thermally driven outflows and the mass-outflow rates.
We compute the energy angular-momentum parameter space for the accretion flows which will
produce such outflows, and its dependence on viscosity parameter. In numerical simulation with
Shakura \& Sunyaev viscosity prescription,
shock location seems to increase with the increase of viscosity parameter \citep{lmc98,ldc11}, while 
theoretical investigations with Chakrabarti-Molteni viscosity prescription showed the opposite
phenomenon \citep{cd07,dc08}. We address this issue and attempt to remove any ambiguity.
Since the post-shock disc could well be the elusive Compton
cloud, while the Shakura-Sunyaev viscosity might be more physical since it satisfies
proper inner boundary conditions around black holes \citep{bl03}, studies of shock
driven outflows and its dependence on viscosity would throw light in understanding
the radio X-ray correlation from X-ray binaries, especially, whether the jet becomes stronger or weaker
with the increasing viscosity parameter,
this would have an interesting connotation in interpreting observations. Since the post-shock disc can produce
high energy photons, then some part of the thermal energy gained through shocks will be dissipated as radiations.
We study the issue of origin of such outflows in presence of dissipative shocks too.

In the next section, we present the simplifying assumptions and equations of motion. In section 3,
we present the methodology of solution. In section 4, we present the solutions, and in the last section
we present discussion and concluding remarks.

%%%%%%%%%%%%%%%%SECTION 2 %%%%%%%%%%%%%%%%%%%%%%%%%%%%%%%%%%%%%%%%%%%%%%
\begin{figure}
\begin{center}
\epsfig{figure=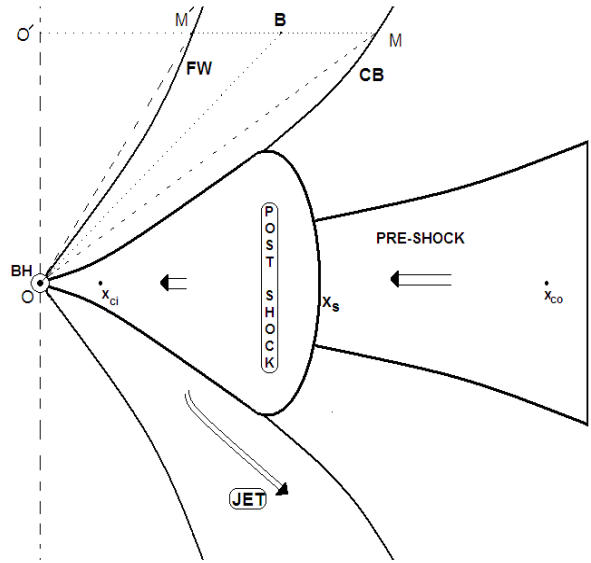,height=8.cm,width=8.cm,angle=0} 
\end{center}
\caption{Schematic diagram of accretion-ejection flow geometry. The black hole (BH) with its horizon
is shown at $O$. $OO^{\prime}$ is the $z$ axis. $r_{CB}=OM$ and $r_{FW}=OM^{\prime}$ are spherical
radial distances
of the Centrifugal Barrier (CB) and Funnel Wall (FW). Moreover, $x_{FW}=O^{\prime}M^{\prime}$,
$x_{CB}=O^{\prime}M$ are indicated. The two physical critical points $x_{ci}$ and
$x_{co}$ of the accretion disc, and the shock location $x_s$ is also shown.
The jet flows through CB and FW.}
\label{lab:ims_fig}
\end{figure}
\section{Model Assumptions and Equations of motion}
It has recently been stressed that black hole rotation plays no, or little part
in generating or powering these jets \citep{fgr10}. So it is expected that
plasma processes or fluid properties of the disc would generate jets.
In this paper, we consider a non-rotating black hole
and focus only on the fluid properties of accretion disc, which may be considered to be
responsible for jet generation.
We have assumed the axis-symmetric disc-jet system to be in steady state.
The space-time properties around a Schwarzschild black hole is described by the pseudo-Newtonian
potential introduced by \citet{pw80}. 
The viscosity prescription in the disc is described by the Shakura-Sunyaev prescription.
We ignore any cooling mechanism in order to focus on the effect of viscosity.
The jets are tenuous and should have less differential rotation than the accretion disc,
as a result the viscosity in jets can be ignored. To negate any resulting torque, the
angular momentum at the jet base is assumed same as that of the local value of angular
momentum of the disc.
The accretion disc occupies the space on or about the equatorial plane. But the jet
flow geometry is described about the axis of symmetry. We first
present the equations of motions of the accretion disc and the jet
separately in the subsequent
part of this section, and then in the next section, we describe the method to obtain the self consistent
accretion-ejection solution. 
In this paper we have used the geometric unit system where, $2G=M=c=1$ ($M$ is the mass of the black
hole, $G$ is the Gravitational constant). Therefore, in this representation the units of length,
mass and time are the Schwarzschild radius or $r_g=2GM/c^2$, $M$, and
$r_g/c=2GM/c^3$, respectively, consequently the unit of speed is $c$.

\subsection{Equations of motion for accretion} 
\label{subsec:hbeta}

The equations of motion for viscous accreting matter around the equatorial plane,
in cylindrical coordinates ($x$, $\phi$, $z$) are given by,

\noindent the radial momentum equation:
%\begin{center}
\begin{equation}
u\frac{du}{dx} + \frac{1}{\rho}\frac{dp}{dx} + \frac{1}{2(x-1)^2}-\frac{\lambda^{2}(x)}{x^{3}} = 0
\label{rme.eq}
\end{equation}
%\end{center}
The mass-accretion rate equation:
%\begin{center}
\begin{equation}
\dot{M}=2\pi\Sigma u x,
\label{mf.eq}
\end{equation}
The mass accretion rate ${\dot M}$ is a constant of motion, except at the regions from where the mass
may be lost into the jets. We present the exact form of the conservation of ${\dot M}$ later in the paper.
%\end{center}
The angular momentum distribution equation:
\begin{equation}
u\frac{d\lambda(x)}{dx}+\frac{1}{\Sigma x}\frac{d(x^{2}W_{x\phi})}{dx}=0
\label{amde.eq}
\end{equation}
The entropy generation equation:
%\begin{center}
\begin{equation}
\Sigma u T\frac{ds}{dx}=Q^{+}-Q^{-}.
\label{ege.eq}
\end{equation}
The local variables $u,~a,~p,~\rho$ and $\lambda$ in the above equations are the radial bulk velocity,
sound speed, isotropic pressure and specific angular momentum of the flow, respectively.
Here, $\Sigma=2\rho h$ and $W_{x\phi}$ are the vertically integrated density and the viscous stress tensor
\citep{mkfo84}. Other quantities  like the the entropy density, the local
temperature, and the local half height of the disc are given by, $s$, $T$, and $h$,
respectively. The local heat gained and
lost by the flow are given by $Q^{+}=W_{x\phi}^{2}/({\eta})$ and $Q^{-}$.

The constant of motion of the flow is obtained, by
integrating equation (\ref{rme.eq}) with the help of  equations (\ref{mf.eq} --- \ref{ege.eq}),
we find that the energy per unit mass is given by
\begin{equation}
E =\frac{u^{2}}{2} + \frac{a^{2}}{\gamma-1}-\frac{\lambda^{2}}{2x^{2}}+\frac{\lambda_{0}\lambda}{x^{2}}
+\Phi(x),
\label{ge.eq}
\end{equation}
and is also called as the specific grand energy of the flow, and is conserved throughout the flow even
in the presence of viscous dissipation \citep{gl04}, except across a dissipative shock (see, section 3.1.1).
In Eq. (\ref{ge.eq}), $\Phi(x)=-0.5/(x-1)$ is the pseudo-Newtonian gravitational potential.

The viscous stress is given by
\begin{equation}
W_{x \phi}= \eta x\frac{d\Omega}{dx},
\label{strs.eq}
\end{equation}
where, 
$\eta=\rho\nu h$ is the dynamic viscosity coefficient, 
$\nu=\alpha a^{2}/(\gamma \Omega_{k})$ is the kinematic viscosity, $\alpha$ is the
Shakura-Sunyaev viscosity parameter, $\Omega$ and $\Omega_{k}$ are
the local angular velocity and local Keplerian angular velocity, respectively.
Considering hydrostatic equilibrium in vertical direction, the local disc half height is obtained as:
\begin{equation}
h(x)=\sqrt{\frac{2}{\gamma}}ax^{1/2}(x-1)
\label{hh.eq}
\end{equation}
The adiabatic sound speed is defined as 
\begin{equation}
a=\sqrt{\frac{\gamma p}{\rho}}
\label{as.eq}
\end{equation}
where $\gamma$ is adiabatic index.
The expression of the entropy-accretion rate is given by, 
\begin{equation}
\dot{\cal{M}}(x)=\frac{a^{(2n+1)}ux}{\Omega_{k}},
\label{ear.eq}
\end{equation}
If there is no viscosity \ie  $\alpha=0$, then $\dot{\cal{M}}$ is constant except
at the shock. At shock the entropy accretion rate will suffer discontinuous jump. The
immediate pre-shock and post shock entropy-accretion rate
denoted as ${\dot {\cal M}}_-$ and ${\dot {\cal M}}_+$, are related as ${\dot {\cal M}}_+>{\dot {\cal M}}_-$.
But for $0<\alpha<1$
or viscous flow,
$\dot{\cal{M}}$
varies continuously in the disc since viscosity dissipates and increases the entropy.
If shock exists in viscous flow then, similar to the inviscid case ${\dot {\cal M}}_+>{\dot {\cal M}}_-$.

The gradient of the angular velocity can be obtained by
integrating equation (\ref{amde.eq}) and also by utilizing equation (\ref{mf.eq}) and the expression
of $W_{x\phi}$,
\begin{equation}
\frac{d\Omega}{dx}=- \frac{\gamma u\Omega_{k}(\lambda-\lambda_{0})}{\alpha a^{2}x^{2}}.
\label{daf.eq}
\end{equation}
where, $\lambda_{0}$ is specific angular momentum at the horizon obtained by considering
vanishing torque at the event horizon \citep{w72,bdl08}.
Since $\lambda=x^{2}\Omega$, the radial derivative of $\lambda$ is given by
\begin{equation}
\frac{d\lambda}{dx}=2x\Omega+x^{2}\frac{d\Omega}{dx}.
\label{dsam.eq}
\end{equation}
Moreover, $\Omega_{k}$ denotes the Keplerian angular velocity and defined as
\begin{equation}
\Omega_{k}^{2}(x)=\frac{1}{2x(x-1)^{2}},
\label{Kaf.eq}
\end{equation}
The Keplerian specific angular momentum is defined as
\begin{equation}
 \lambda_k=\Omega_k~x^2=\left[\frac{x^3}{2(x-1)^2}\right]^{1/2}.
\label{Kang.eq}
\end{equation}

Manipulating equations (\ref{rme.eq} --- \ref{ege.eq}), with the help of equations(\ref{hh.eq} ---
\ref{Kang.eq}) we obtain,
\begin{equation}
\frac{du}{dx}=\frac{N}{D}.
\label{du.eq}
\end{equation}
where, 
$$
N=\frac{2}{\gamma+1}\frac{(5x-3)u}{2x(x-1)}+\frac{(\lambda^{2}-\lambda_{k}^{2})u}{a^{2}x^{3}}
+\gamma^{2}\left(\frac{\gamma-1}{\gamma+1}\right)\frac{u^{2}\lambda_{k}(\lambda-\lambda_{0})^{2}}
{\alpha a^{4}x^{4}}
$$ and
$$
D=\frac{u^{2}}{a^{2}}-\frac{2}{\gamma+1}
$$
The gradient of the sound speed is,
\begin{equation}
\frac{da}{dx}=\left(\frac{a}{u}-\frac{\gamma u}{a}\right)\frac{du}{dx}+\frac{(5x-3)a}{2x(x-1)}
+\frac{\gamma (\lambda^{2}-\lambda_{k}^{2})}{ax^{3}}.
\label{da.eq}
\end{equation}

Therefore, the accretion disc problem in vertical equilibrium is solved by integrating Eqs.
(\ref{dsam.eq}, \ref{du.eq}, \ref{da.eq}). 

\subsubsection{Critical point conditions for accretion} 
At large distances away from the horizon the
inward velocity is very small and therefore the flow is subsonic, but, matter enters the black hole with
the speed of light and therefore it is supersonic close to the horizon.
Hence accreting matter around black holes must be
transonic, since it makes a transition from subsonic to supersonic. Therefore, at some location the denominator $D$ of Eq. (\ref{du.eq}), will go to zero,
and hence the numerator $N$ goes to zero too. Such a location is called the sonic point or critical point. 
The critical point conditions are given as:
\begin{equation}
M_{c}^{2}=\frac{u_{c}^{2}}{a_{c}^{2}}=\frac{2}{\gamma+1}
\label{mc.eq}
\end{equation} 
\begin{equation}
\begin{split}
\left[\frac{(5x_{c}-3)M_{c}^{2}}{2x_{c}(x_{c}-1)} \right]a_{c}^{3}+\left[\frac{(\lambda_{c}^{2}-\lambda_{k}^{2})}{x_{c}^{3}}\right]a_{c} \\ +\gamma^{2}\left(\frac{\gamma-1}{\gamma+1} \right)\frac{M_{c}\lambda_{k}(\lambda_{c}-\lambda_{0})^{2}}{\alpha x_{c}^{4}}=0
\end{split}
\label{nc.eq}
\end{equation} 

where $M_{c}, u_{c}, a_{c}, x_{c}$ and $\lambda_{c}$ are Mach number, the bulk velocity, the sound speed,
the radial distance and the specific angular momentum at the critical point, respectively.

The radial velocity gradient at the critical point is calculated by employing the l$^{\prime}$Hospital
rule. 
\begin{equation}
 \left(\frac{du}{dx}\right)_c=\left(\frac{dN/dx}{dD/dx}\right)_{r=r_c}
\label{duc.eq}
\end{equation}
and by combining Eqs. (\ref{da.eq} \& \ref{duc.eq}) we get,
\begin{equation}
\left(\frac{da}{dx}\right)_c=\left(\frac{a_c}{u_c}-\frac{\gamma u_c}{a_c}\right)\left(\frac{du}{dx}\right)_c+\frac{(5x_c-3)a_c}{2x_c(x_c-1)}+\frac{\gamma (\lambda^{2}_c-\lambda_{kc}^{2})}{a_cx^3_c}.
\label{dac.eq}
\end{equation}

So, the solution of Eqs. (\ref{dsam.eq}, \ref{du.eq}, \ref{da.eq}), can only be obtained
if we know the sonic point and its conditions (Eqs. \ref{mc.eq}---\ref{dac.eq}).
 
\subsection{Equations of motion for outflows}

The flow geometry for accretion is described about the equatorial plane, however, the jet or outflow
geometry is about the axis of symmetry. If the outflow posses some angular momentum, then
the outflow geometry should be hollow. Indeed,
numerical simulations by \citet{mrc96} suggests that the outflowing matter tends to emerge out between
two surfaces namely, the funnel wall(FW) and centrifugal barrier(CB).
In Fig.1, the schematic diagram of the jet geometry is shown.
The centrifugal barrier(CB) surface is defined as the pressure maxima surface and is expressed as
\begin{equation}
x_{CB}=[2 \lambda^{2}r_{CB}(r_{CB}-1)^{2}]^{\frac{1}{4}} 
\label{cb.eq}
\end{equation}
where, $r_{CB}=\sqrt{x_{CB}^{2}+y_{CB}^{2}}$, spherical radius of CB.
Here, $x_{CB}$ and $y_{CB}$ are the cylindrical radius and axial coordinate (i.e.,
height at $r_{CB}$) of CB. We compute the jet geometry with respect to $y_{CB}$ i.e.,
$y_{FW}=y_{j}=y_{CB}$, where $y_{FW}$ and $y_{j}$ are the height of FW and the jet at
$r_{CB}$, respectively. The FW is obtained by considering null effective potential and
is given by
\begin{equation}
x_{FW}^{2}=\lambda^{2} \frac{(\lambda^{2}-2)+\sqrt{(\lambda^{2}-2)-4(1-y_{CB}^{2})}}{2} 
\label{fw.eq}
\end{equation}
where, $x_{FW}$ is the cylindrical radius of FW. We define the cylindrical radius of the outflow
\begin{equation}
x_{j}=\frac{x_{FW}+x_{CB}}{2} 
\label{jcr.eq}
\end{equation}
The spherical radius of the jet is given by $r_{j}=\sqrt{x_{j}^{2}+y_{j}^{2}}$. In Fig.1, OB($=r_{j}$)
defines the streamline (solid) of the outflow. The total area function of the jet is defined as,
\begin{equation}
{\cal{A}}=\frac{2 \pi (x_{CB}^{2}-x_{FW}^{2})}{\sqrt{1+(dx_j/dy_j)^2}},
\label{ja.eq}
\end{equation}
where, the denominator is the projection effect of the jet streamline on its cross-section.
The integrated radial momentum equation for jet is given by,
\begin{equation}
{\cal{E}}_{j}=\frac{1}{2}v_{j}^{2}+n a_{j}^{2}+\frac{\lambda_{j}^{2}}{2x_{j}^{2}}-\frac{1}{2(r_{j}-1)}
\label{jse.eq}
\end{equation}
where, ${\cal{E}}_{j}$ is the specific energy, $\lambda_{j}$ is the angular momentum of the jet,
and $n=1/(\gamma -1)$ is the polytropic index. The integrated continuity equation is,
\begin{equation}
\dot{M}_{out}=\rho_{j}v_{j} \cal{A} 
\label{jmf.eq}
\end{equation}
and the entropy generation equation is integrated to obtain the polytropic equation
$(p_{j}=K_{j}\rho_{j}^{\gamma})$ of state for the jet. 
The entropy accretion rate for the jet is given by
\begin{equation}
 {\dot {\cal M}}_j=a^{2n}_jv_j{\cal A}
\label{entroj.eq}
\end{equation}
In Eqs. (\ref{jse.eq}-\ref{jmf.eq}),
the suffix `j' indicates jet variables, where $v_{j}$, $a_{j}$, and $\rho_{j}$ are the velocity,
the sound speed and the density of the jet.

Equations~(\ref{jse.eq}-\ref{jmf.eq} ) are differentiated with respect to $r(=r_{CB})$,
to obtain,
\begin{equation} 
\frac{dv_{j}}{dr}=\frac{\cal N}{\cal D},
\label{gvjet.eq}
\end{equation}
where,
\begin{equation}
{\cal N}=\frac{1}{2(r_{j}-1)^{2}}\frac{dr_{j}}{dr}
-\frac{\lambda_{j}^{2}}{x_{j}^{3}}\frac{dx_{j}}{dr}
-\frac{a_{j}^{2}}{{\cal{A}}}\frac{d{\cal{A}}}{dr},
\end{equation}
and,
\begin{equation}
 {\cal D}=\frac{a_{j}^{2}}{v_{j}}-v_{j}.
\end{equation}

The critical point$(r_{jc})$ condition for jet is,
\begin{equation}
v_{jc}^{2}=a_{jc}^{2}=\left[\frac{1}{2(r_{jc}-1)^{2}}\left(\frac{dr_{j}}{dr}\right)_{r_{c}}
-\frac{\lambda_{jc}^{2}}{x_{jc}^{3}}\left(\frac{dx_{j}}{dr}\right)_{r_{c}}  \right] \left[\frac{1}{{\cal{A}}_{c}}\left(\frac{d{\cal{A}}}{dr}\right)_{r_{c}}  \right]^{-1} .
\label{jetc.eq}
\end{equation}

The outflow solution is obtained by integrating Eqs. (\ref{gvjet.eq}), with the help of
Eq. (\ref{jetc.eq}). The outflow equations can be determined uniquely if ${\cal{E}}_{j}$ and $\lambda_{j}$
are known, because, the value of ${\cal{E}}_{j}$ and $\lambda_{j}$ determines $r_{jc}$.
However, for consistent accretion-ejection solution, both accretion and jet part have to be solved
simultaneously, a technique previously used by \citet{cd07}, and has been refined in this paper. 

\section{Methodology}

The accretion-ejection solutions are self consistently and simultaneously solved,
and we present the methodology to find such solutions in this section.
We start with the solution of the accretion disc.
It has been mentioned before that Eqs. (\ref{dsam.eq}, \ref{du.eq}, \ref{da.eq})
can be integrated, if we know the sonic point $x_c$. 
One of the long standing problem in accretion physics is to determine the sonic point of the flow
in presence of a viscous stress of the form Eq. (\ref{strs.eq}), which keeps the angular momentum
equation in a differential form [Eq. (\ref{dsam.eq})] rather than an algebraic form \citep{c96,gl04}.
The problem is compounded by the fact that although quantities on the horizon are
known, but the coordinate singularity on the horizon makes it
difficult to solve the equations by taking those as the starting values.
The problem could be circumvented if the asymptotic values of the
flow variables are known close to the horizon.

With a clever use of conservation equations \citet{bl03} found the asymptotic distribution of
the specific angular momentum and the radial velocity close to the horizon, and they are,
 
\begin{equation}
\lambda(x)=\lambda_{0}\left[1+\frac{2\alpha}{\gamma r_{g}}\left(\frac{2}{r_{g}}\right)^{1/2}\left(\frac{{\dot{\cal{M}}}^{2}}{2r^3_g}  \right)^{\frac{\gamma-1}{\gamma+1}}(x-r_g)^{\frac{\gamma+5}{2\gamma+2}}\right], \ x\rightarrow 1 %r_{g}
\label{asam.eq}
\end{equation}
and
\begin{equation}
u(x)=u_{ff}(x)\left[1+\frac{2Ex^{2}-\lambda_{0}^{2}-(\gamma+1)f(x)}{x^{2}u_{ff}^{2}(x)-(\gamma-1)f(x)}  \right]^{1/2}, \ x\rightarrow 1 %r_{g}
\label{au.eq}
\end{equation}
where the function $f(x)$ is $f(x)=\frac{2x^{2}}{\gamma^{2}-1}\left[\frac{{\dot{\cal{M}}}^{2}}{2x^{3}(x-r_g)} \right]^{\frac{\gamma-1}{\gamma+1}}$ and the free fall velocity in the pseudo-Newtonian potential geometry is given by
\begin{equation}
u_{ff}(x)=\frac{1}{\sqrt{(x-1)}}
\label{uff.eq}
\end{equation}

\begin{figure}
\begin{center}
\epsfig{figure=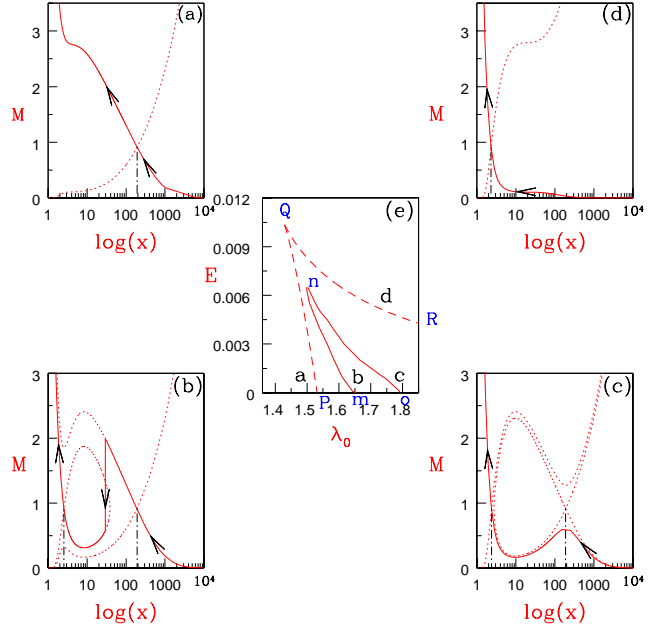,height=8.5cm,width=8.5cm,angle=0}
\end{center}
\caption{The domain for multiple critical point $MCP$ (dashed) and shock (solid) in $E-\lambda$ parameter space (e)
is presented. The solution topologies or the plot of Mach number $M$ with $log(x)$ 
of the O type for $E=0.001,~\lambda=1.5$ (a); A type for $E=0.001,~\lambda=1.68$ (b); W type fort
$E=0.001,~\lambda=1.75$ (c); and I type for $E=0.005,~\lambda=1.77$(d), are presented. The solution type and their location in the parameter space is also indicated in (e). All the plots are
for $\alpha=0$. The dotted lines in the panels named a, b, c, and d show all possible solutions while, the solid
line with arrow heads show the
actual accretion solution. The vertical long-short dashed line shows the location of the sonic points.}
\label{lab:ims_fig}
\end{figure}

\subsection{Integration procedure of accretion solution}
A position very close to the horizon $x=x_{in}=1.01$ is chosen. 
For given values of the parameters $\alpha$, $E$, $\lambda_{0}$ at the horizon and $\dot{\cal{M}}_{in}$ at
$x_{in}$,
Eqs. (\ref{asam.eq}), (\ref{au.eq}) and (\ref{uff.eq}) can be used to determine the asymptotic values of the fluid variables
close to the horizon, and which can be used as the initial values 
for the integration. With these values, Eqs. (\ref{dsam.eq}, \ref{du.eq}, \ref{da.eq})
are integrated outwards, simultaneously checking for the location of the sonic point by using
Eqs. (\ref{mc.eq}---\ref{dac.eq}). The sonic point $x_c$ is determined iteratively, and once it is found,
the sonic point conditions
are used to integrate the equations of motions, outwards. It is well known that matter with
angular momentum may posses {\it multiple critical points} (MCP). But the flow may pass through various
sonic points only if a shock is present, in other words, the existence of MCP is mandatory
for the formation of shock in black-hole accretion.
Only a supersonic flow can undergo shock transition and so existence of one sonic point at larger
distance (\ie $x_{co}$) from the horizon is warranted. The post shock flow is subsonic. 
However, the inner boundary condition of black hole accretion is supersonic,
and hence the subsonic post-shock flow has to pass through another sonic point ($x_{ci}$),
before it dives into the black hole. In other words, shock in black hole accretion may exist
only if MCP exists. 
It is to be remembered though, that there is no smooth transition between various branches
of the solution passing through different sonic points. This is true for both inviscid and adiabatic flows
as well as viscous and non-adiabatic flow. However, if the flow is inviscid, for a given value of $E$ and $\lambda_{0}$
all possible sonic points are known apriori. Incase of viscous flow which
is following a viscosity prescription of the form Eq. (\ref{strs.eq}), the existence of multiple
sonic points can only be ascertained only if there is a shock.
%% %

\subsubsection{Shock equations}
The Rankine Hugoniot (RH) shock conditions are obtained by conservation of the mass, momentum and energy
flux across the shock. In presence of mass loss and energy loss, the shock conditions are given by,
the modified mass conservation,
\begin{equation}
\dot{M_{+}}=\dot{M_{-}}-\dot{M_{out}}=\dot{M_{-}}(1-R_{\dot{m}}),
\label{mfs.eq}
\end{equation}
This equation effectively divides the accretion rate of pre-shock accretion disc (${\dot M}_-$)
into two channels, namely the post-shock accretion disc (represented by ${\dot M}_+$)
and the jet (represented by ${\dot M}_{out}$).
The modified momentum conservation,
\begin{equation}
W_{+}+\Sigma_{+} u_{+}^{2}=W_{-}+\Sigma_{-} u_{-}^{2},
\label{rmf.eq}
\end{equation}
and the third shock condition is the modified energy conservation
\begin{equation}
E_{+} =E_{-}-\Delta {E},
\label{ef.eq}
\end{equation}
where,  $R_{\dot{m}}$ is the relative mass outflow rate given by,
\begin{equation}
R_{\dot{m}}=\frac{\dot{M_{out}}}{\dot{M_{-}}}.
\label{rmd.eq}
\end{equation}

Here, subscripts minus(-) and plus(+) denote the quantities before and after the shock. $W$ is the vertically integrated pressure.
In absence of massloss ($R_{\dot m}=0$) or dissipation (\ie $\Delta E=0$), Eqs. (\ref{mfs.eq}, \ref{rmf.eq},
\ref{ef.eq}) reduce to the standard RH shock conditions. Since the dissipation is expected at the shock location, it is assumed that energy dissipation takes place mostly through the thermal Comptonization \citep{ct95,mc10} and is likely to be very important within a distance $dx$ inside the shock where the optical depth is around unity. So this energy dissipation in the post shock flow reduces the temperature of the flow and the loss of energy is proportional to the temperature difference between the post-shock and the pre-shock flows, i.e. 
\begin{equation}
\Delta {E} = f_{e} n (a_{+}^{2}-a_{-}^{2}).
\label{dele.eq}
\end{equation}
where $f_{e}$ is the fraction of the difference in thermal energy dissipation across the shock transition and $n$ is the polytropic index. We use $f_e$ as a parameter, in presence of detailed radiative processes
$f_e$ can be self-consistently determined.
Since shock width is infinitesimally thin, so  
we assume that $d\Omega/dx$ is continuous across the shock.
The angular momentum jump condition is calculated by considering the conservation of angular momentum flux,
and is given by
\begin{equation}
\lambda_{-}=\lambda_{+}+C_{sh}\left[\frac{a_{+}^{2}}{u_{+}}-\frac{a_{-}^{2}}{u_{-}}\right],
\label{sams.eq}
\end{equation}
where, $C_{sh}=-u_{+}(\lambda_{+}-\lambda_{0})/a_{+}^{2}$.
Equation (\ref{sams.eq}) can be re-written as,
\begin{center}
 \begin{equation}
  (\lambda_- - \lambda_0) = \frac{a_-^2 u_+(\lambda_+ - \lambda_0)}{a_+^2 u_-}
\label{lsh.eq}
 \end{equation}
\end{center}
Since at shock, $a_+>a_-$ and $u_->u_+$, therefore, $\lambda_+>\lambda_-$.

Using shock condition equations (\ref{mfs.eq}) and (\ref{rmf.eq}), the pre-shock sound speed and bulk velocity can
be written as,
\begin{equation}
a_{-}^{2}=k_{1}u_{-}-\gamma u_{-}^{2}
\label{ass.eq}
\end{equation}
where, $k_1=(a_{+}^{2}+\gamma u_{+}^{2})/(fu_{+})$, $f=1/(1-R_{\dot {m}})$.
Now substituting for $a_{-}$ and $\lambda_{-}$ in equation (\ref{ef.eq}),
we find a quadratic equation of $u_{-}$ as,
\begin{equation}
C_{2}u_{-}^{2}+C_{1}u_{-}+C_{0}=0
\label{us.eq}
\end{equation}
where, 
$$C_{2}=\left[\frac{1}{2}-n\gamma (1+f_{e})-\frac{\gamma^{2}C_{sh}^{2}}{2x_{s}^{2}}\right],
$$
$$C_{1}=\left[n (1+f_{e})k_{1}-\frac{\gamma C C_{sh}}{x_{s}^{2}}+\frac{\gamma
\lambda_{0}C_{sh}}{x_{s}^{2}}\right],
$$
$$
C_{0}=\left[\frac{\lambda_{0}C}{x_{s}^{2}}-\frac{C^{2}}{2x_{s}^{2}}-k_{2}  \right],
$$
$$C=\lambda_{+}+C_{sh}\left[\frac{a_{+}^{2}}{u_{+}}- k_{1} \right]
$$
and
$$
k_{2}=E_{+}-\Phi(x_s)+f_{e}na_{+}^{2}.
$$

In terms of the shock quantities, the mass outflow rate is given by
\begin{equation}
R_{\dot m}={\dot {M}}_{\rm out}/{\dot {M}}_{-}
=\frac{Rv_j(x_s) {\mathcal A}(x_s)}{4 \pi \sqrt{\frac{2}{\gamma}}x^{3/2}_s (x_s-1)a_+u_{-}}
=R{\cal V}{\cal G},
\label{rmdot.eq}
\end{equation}
where, $R=\Sigma_+/\Sigma_-$ is the compression ratio across the shock, ${\cal V}=v_j(x_s)/u_-$
is the ratio of jet base velocity and the pre-shock velocity of the disc ($u_-$), and
${\cal G}={\cal A}(x_s)/[4 \pi (2/\gamma)^{1/2}x^{3/2}_s (x_s-1)a_+]$
is the ratio of the jet cross-sectional area at $r=x_s$ and the post-shock accretion disc
cross-sectional area.

\begin{figure}
\begin{center}
\epsfig{figure=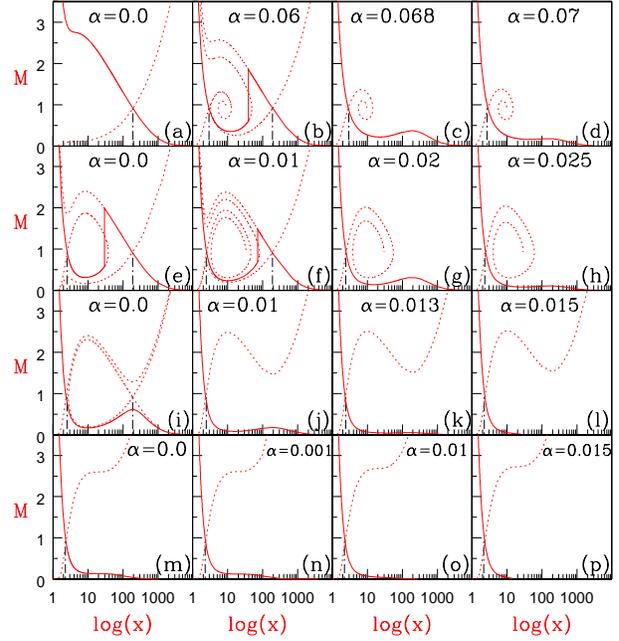,height=9.cm,width=8.7cm,angle=0} 
\end{center}
\caption{Variation of $M$ with $log(x)$ for the accretion solutions with different viscosity parameter $\alpha$.
In a, e, i, m we present inviscid solutions corresponding to the O, A, W, I type of solutions from Figs. 2a-d.
Towards right
$E,~\lambda_0$ is kept the same but
$\alpha$ is increased. The flow parameters for which these plots are generated $E=0.001,~\lambda=1.5$ (a, b, c, d);
$E=0.001,~\lambda=1.68$ (e, f, g, h); $E=0.001,~\lambda=1.75$ (i, j, k, l) and $E=0.005,~\lambda=1.75$ (m, n, o, p).
The viscosity parameter $\alpha$ is mentioned on the figure. The vertical long-short dashed line shows the location of
sonic points.}
\label{lab:eov_fig}
\end{figure}
\begin{figure}
\epsfig{figure=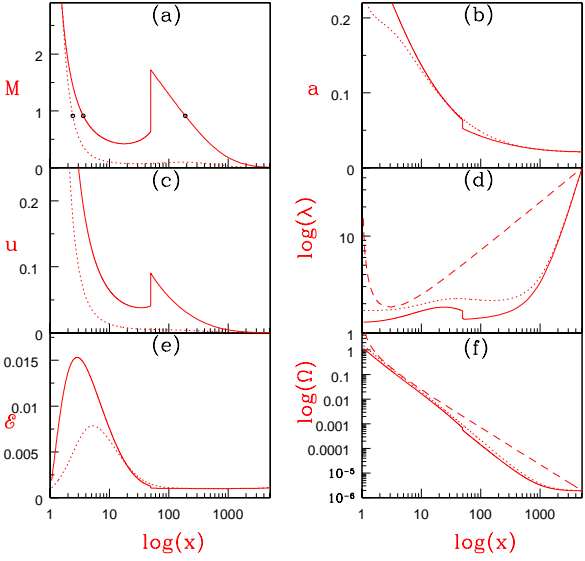,height=8.cm,width=8.cm,angle=0} 
\caption{Comparison of shock free (dotted) and shocked solution (solid).
Both the solution starts with the
same outer boundary condition $x_{inj}=5117$, $\lambda_{inj}=\lambda_K(x_{inj})=50.6$,
$E=10^{-3}$. The viscosity parameter for the shock free solution is $\alpha=0.02$ (dotted)
and that for the shocked one $\alpha=0.15$ (solid). 
Various flow variables are plotted are $M$ (a), $a$ (b), $u$ (c), $\lambda$ (d), ${\cal E}$ (e)
and $\Omega$ (f) as a function of $x$. 
The sonic points are marked with open circles (a).
The shock free solution has only one $x_{ci}$, the shocked solution has both $x_{ci}$ and $x_{co}$
and the shock location is at $x_s=49.68$. Dashed plots of $\lambda_K$ and $\Omega_K$ in (d) and (f) are Keplerian angular momentum and angular velocity
drawn for comparison.}
\label{lab:adsh_fig}
\end{figure}
\subsection{Accretion-Ejection Solution}
The accretion-ejection is computed self-consistently. We have set
$\gamma=1.4$ and $x_{in}=1.01$ throughout the paper.
The method to find the accretion-ejection solution is as follows,
\begin{enumerate}
 \item First we assume that $R_{\dot m}=0$. With chosen values of $\lambda_0$, $E$, $\alpha$
and $f_e$, we follow the procedure described in section 3.1, \ie
determine the inner sonic point iteratively and integrate outwards. 
Equations (\ref{sams.eq}, \ref{ass.eq}, \ref{us.eq}) are checked to calculate
the pre-shock quantities. We find out the outer sonic point ($x_{co}$) iteratively, from the pre-shock
quantities.
Two possibilities may arise, either the flow passes through only one sonic point and gives a smooth
solution, or, finds a stable shock solution and passed through two sonic points.
The location of the jump for which $x_{co}$ exists is the virtual shock
location (${\tilde x}_s$).
\item  Once ${\tilde x}_s$ is found out, we assign ${\cal E}_j={\cal E}_+$
and $\lambda_j=\lambda_+$ and solve jet Eqs. (\ref{gvjet.eq} \& \ref{jetc.eq})
and compute the corresponding $R_{\dot m}$.
\item We put this value of $R_{\dot m}$ into Eqs. (\ref{sams.eq}-\ref{us.eq})
to find a new shock location .  
\item Steps (ii) \& (iii) are repeated till the temporary shock location ${\tilde x}_s$
converges to the actual shock location ($x_s$).
The converged shock solution therefore gives the actual jet solution too. We find that
the ${\dot {\cal M}}_j>{\dot {\cal M}}_-$ and
${\dot {\cal M}}_+>{\dot {\cal M}}_-$. Since matter prefers higher entropy
solutions, therefore, the post-shock fluid would prefer both the channels, one through the $x_{ci}$  onto the black hole, and the other through 
$r_{jc}$ and out of the disc-jet system. In other words, a shock in accretion would generate
a bipolar outflow from the post-shock region too.
\end{enumerate}
The outer boundary is chosen as $x_{inj}=10^4r_g$, or the distance at which
$\lambda(x_{inj})=\lambda_K(x_{inj})$, which ever is shorter.
\begin{figure}
\begin{center}
\epsfig{figure=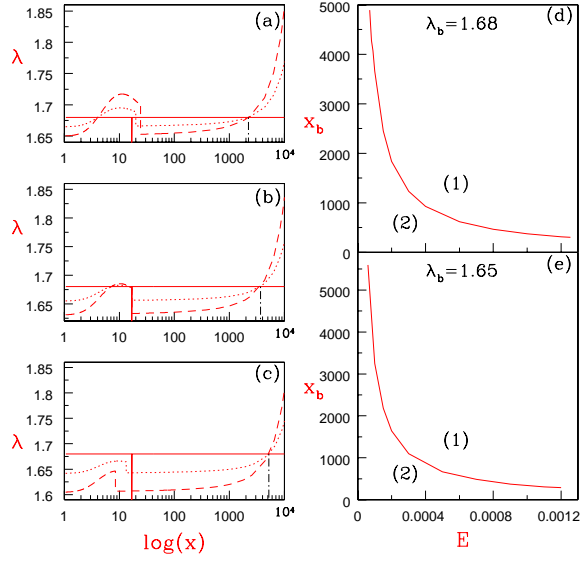,height=8.cm,width=8.cm,angle=0} 
\end{center}
\caption{Variation of $\lambda$ with $log(x)$ of shocked accretion flow (a, b, c),
for parameters
$ E=0.0001,~\&~\lambda_{b}=1.68$. Each curve
represents $\alpha=0.0$ (solid), $\alpha=0.0075$(dotted) and $\alpha=0.015$ (dashed).
First three panels are for $x_{b}=2220 r_{g}$(a), $x_{b}=3660 r_{g}$ (b), and $x_{b}=5230 r_{g}$ (c).
Vertical solid line and dash-dotted line show the shock location for $\alpha=0.0$ curve and outer boundary location $(x_{b})$, respectively.
The variation of limiting $x_{b}$ with $ E$ for $\lambda_{b}=1.68$ (d), and $\lambda_{b}=1.65$ (e). Domain 1 represents all $x_{b}$ for which $x_s$ decrease with the increase with $\alpha$, but for any $x_{b}$
in 2, $x_s$ will increase with the increase of $\alpha$.}
\label{lab:adaf_fig}
\end{figure}
%%%%%%%%%%%%%%%%%%%SECTION 3 %%%%%%%%%%%%%%%%%%
\section{Solutions}
We present every possible way matter can dive into the black hole. In this section,
we start with 
accretion solutions without considering massloss or dissipation at the shock front
and study the effect of viscosity on accretion solution.
Then we present the accretion-ejection solution and study how the viscosity
can affect the mass outflow rates. And finally we present accretion-ejection solutions
in presence of dissipative shocks and show the effect of both the viscosity and 
dissipation at the shock front.

\subsection{All possible accretion solutions in the advective regime}

Since black hole accretion is necessarily transonic, at first, we present the simplest rotating
transonic solutions \ie inviscid global solutions
(global solution$\equiv$ which connect horizon and large distances) in Figs. \ref{lab:ims_fig}a-d. 
The inviscid solutions are presented as an attempt to recall the simplest accretion case \citep{c89,dcc01}.
For inviscid
flow the constants of motions are $E$, $\lambda~(=\lambda_0)$, and if outflows are not present
then, ${\dot M}$ is also a constant of motion.
Moreover, in absence of viscosity, it straightaway follows from Eq. (\ref{ge.eq}) that
$E={\cal E}=u^2/2+na^2+\lambda^2/(2x^2)-0.5/(x-1)$, where ${\cal E}$
is the Bernoulli
parameter or the specific energy of the flow. In Figs. \ref{lab:ims_fig}a-d, accretion
solutions in terms of the Mach number $M~(=u/a)$ distribution are presented, and
in Fig. \ref{lab:ims_fig}e, the $E-\lambda_0$ parameter space for multiple sonic point
and the shock is presented too. Depending on $E$ \& $\lambda_0$ of the flow,
the solutions are also different. If the $\lambda_0$ is low, there is only one outer sonic point $x_{co}$,
and solution type is O-type or Bondi type ($E=0.001,~\lambda_0=1.5$, Fig. \ref{lab:ims_fig}a).
As $\lambda_0$ is increased, the number of physical 
sonic points increases to two and the accretion flow which becomes supersonic through $x_{co}$
can enter the black hole through $x_{ci}$ if a shock condition is satisfied. Although the shock
free solution is possible but in this part of the parameter space a shocked solution will be preferred
because a shocked solution is of higher entropy (or in other words of higher ${\dot {\cal M}}$). Such a class
of solution is called A-type ($E=0.001,~\lambda_0=1.68$, Fig. \ref{lab:ims_fig}b).
For even higher $\lambda_0$ only one sonic point is possible
($E=0.001,~\lambda_0=1.75$ for Type W; and $E=0.005,~\lambda_0=1.77$ for I shown in Figs \ref{lab:ims_fig}c and d),
and the solutions are wholly subsonic till $x_{ci}$ and then dives on to blackhole supersonically. W type solutions are different than I type, in the sense, W type is still
within the MCP domain while I type is not. Moreover, I type is a smooth monotonic solution, although W is smooth
and shock free but is not monotonic and has an extremum at around $x_{co}$.
The parameter space $E-\lambda_0$, bounded by solid line (mno) shows the RH shock parameter space,
while the dotted one (PQR) shows the MCP domain (Fig. 2e). It is to be noted, that in the inviscid limit, accretion
is only possible if $\lambda_0 < \lambda_K$.

\begin{figure}
\epsfig{figure=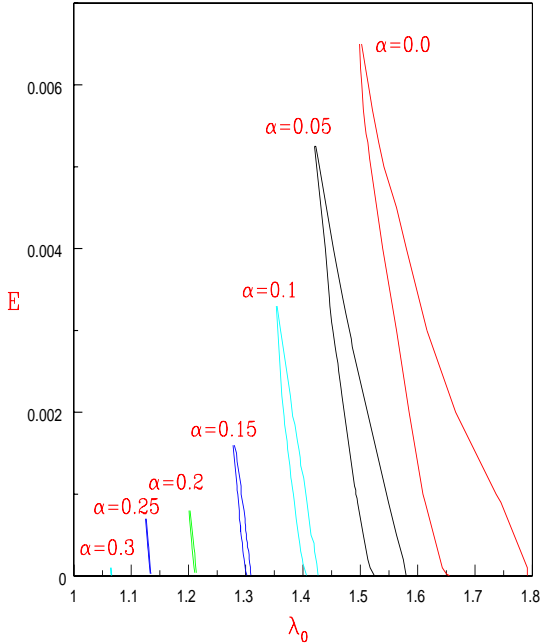,height=10.cm,width=8.cm,angle=0} 
\caption{$E$---$\lambda_0$ parameter space for shock, for various
viscosity parameters $\alpha=0$, $0.05$, $0.1$, $0.15$, $0.2$,
$0.25$, $0.3$ marked on the figure. No energy dissipation ($f_{e}=0.0$) at the shock 
and no mass loss ($f=1.0$) is assumed.}
\label{lab:sokparam_fig}
\end{figure}

Figures \ref{lab:eov_fig}a, \ref{lab:eov_fig}e, \ref{lab:eov_fig}i, \ref{lab:eov_fig}m, represent
the inviscid solutions corresponding to the O type solutions ($E,~\lambda_0=~
0.001,~1.5$), A type ($E,~\lambda_0=~0.001,~1.68$), W-type ($E,~\lambda_0=~0.001,~1.8$),
and I-type ($E,~\lambda_0=~0.006,~1.8$), and are also depicted in Figs. 2a-d. Keeping $E$ \& $\lambda_0$ same,
we increase the viscosity parameter in the right direction. Figures \ref{lab:eov_fig}a-d, has same $E~\&~\lambda_0$,
but progressively increasing $\alpha=0.06$ (Fig. \ref{lab:eov_fig}b), $0.068$ (Fig. \ref{lab:eov_fig}c) and $0.1$
(Fig. \ref{lab:eov_fig}d).
Similarly for Figs. \ref{lab:eov_fig}e-h, $E-\lambda_0$ is same but different $\alpha$, so is the case for
Figs. \ref{lab:eov_fig}i-l,
and Figs. \ref{lab:eov_fig}m-p.
Interestingly, the viscous I type is in principle the much vaunted ADAF type solution presented in Figs 3d, 3h, 3k-l,
and 3n-p. 
The shock-free solution is characterized by monotonic spatial distribution of flow variables, and wholly subsonic
except very close to the horizon which are essentially the viscous I type solutions
and has also been shown by \citet{lgy99,bdl08,dbl09}.
It is evident from Figs. \ref{lab:eov_fig}a-p, that the effect of viscosity is to create additional
sonic points in some part of the parameter space, opening up of closed topologies, and might trigger shock
formation where there was no shock, while removing both shock and multiple critical points in other regime
of the parameter space.
All of this is achieved by removing angular momentum outwards while increasing the entropy inwards. 
In this connection one may find two kind of critical viscosity parameters in the advective domain.
If the inviscid solution is
O type (Fig. 3a), then there would be a lower bound of critical viscosity $\alpha_{\rm cl}$
which would transport angular momentum in a manner that would trigger the standing shock.
And there would be another upper bound of viscosity parameter $\alpha_{\rm cu}$ that would quench the standing shock.
While, if
the inviscid solution has a shock to start with (Fig. 3e), then there could only be $\alpha_{\rm cu}$.
For the case presented in Figs. 3a-d, $\alpha_{\rm cl}=0.0465$ and $\alpha_{\rm cu}=0.065$.
And for the case presented in Figs. 3e-h, $\alpha_{\rm cu}=0.0126$.

So far, we have compared solutions
with different viscosity but same internal boundary condition ($E,~\lambda_0$). Let us compare two
solutions with same outer boundary condition.
In Figs. \ref{lab:adsh_fig}a-e, we compare two solutions starting at the same outer boundary
$x_{inj}=5117$, same grand energy $E=10^{-3}$ and $\lambda_{inj}=\lambda_K(x_{inj})=50.6$.
Both the solution starts with the same entropy $\sim {\dot {\cal M}}_{inj}=2.72\times 10^{-5}$.
For $\alpha=0.15$ (solid)
the accretion solution
has a standing shock at $x_s=49.68$ (the vertical jump in solid),
while the shock free solution is for $\alpha=0.02$ (dotted).
The flow variables plotted are $M$ (Fig. \ref{lab:adsh_fig}a), $a$ (Fig. \ref{lab:adsh_fig}b),
$u$ (Fig. \ref{lab:adsh_fig}c), $\lambda$ (Fig. \ref{lab:adsh_fig}d), and ${\cal E}$
(Fig. \ref{lab:adsh_fig}e).
Incase of inviscid flow $E={\cal E}$ is constant, for viscous flow $E$ is still a constant of motion
but ${\cal E}$, the specific energy, varies with $x$ as is shown in Fig. \ref{lab:adsh_fig}e.
The inner part of the
shocked solution is faster, hotter, and of lesser angular momentum. The
inner boundary for the shocked flow is $E=10^{-3},~\lambda_0=1.29$, while that for the shock free
%1.287
solution is $E=10^{-3},~\lambda_0=1.7$.
The jump in $\lambda$ at the shock follows Eq. \ref{lsh.eq}.
Interestingly, the post shock flow is of higher entropy [${\dot {\cal M}}(x\sim 1)=7.49\times 10^{-5}$] than 
the inner part of the shock free solution [${\dot {\cal M}}(x\sim 1)=7.12\times 10^{-5}$].
Although it may seem contradictory  that a shock free solution exists for lower $\alpha$,
while shocked solution appears for higher $\alpha$,
however, it is to be remembered that with 
such high $\lambda_{inj}$, there would be no accretion solution for $\alpha=0$.
One has to have a certain non-zero viscosity to even have a global (that connects horizon and outer edge)
solution.
Infact, for this particular case we have identified a limiting viscosity parameter $\alpha_1=0.00747$,
such that advective global solutions are possible for any viscosity parameter $\alpha \geq \alpha_1$.
And for flows starting with such high initial $\lambda$, if the viscosity parameter $\alpha$ is small, then
the $\lambda$
distribution will be higher. Therefore, for such high $\lambda$ radial velocities will not be high enough to
become supersonic and form a shock.
So one needs higher $\alpha \geq \alpha_{\rm cl}=0.1496$ to reduce the angular momentum to the extent that
may produce standing accretion shock. And if $\alpha$ is increased beyond another critical
value $\alpha_{\rm cu}=0.1555$, steady shock is not found.
Hence one can identify three critical $\alpha$'s for
accretion flows starting with the same outer boundary condition, where the angular momentum
at the outer boundary is local Keplerian value. Interestingly, $\alpha$ prescription originally
invoked to generate Keplerian disc \citep{ss73}, can produce sub-Keplerian accretion flow with or without shock
even if angular momentum is Keplerian at the outer boundary. This is not surprising since,
the gravitational
energy released by the infalling matter, would be converted to kinetic energy and
thermal energy (by compression and viscous dissipation). If the gravitational
energy is converted to thermal energy and only the rotation part of kinetic energy, and also if
the thermal energy gained by viscous dissipation is efficiently radiated away, then one would produce
Keplerian disc solutions. The $\lambda_K$ (Fig. \ref{lab:adsh_fig}d) and $\Omega_K$
(Fig. \ref{lab:adsh_fig}f) are presented for comparison. In the present paper, the radiative processes have
been ignored which produces hot flow, but the advection terms have not been ignored. Therefore,
sub-Keplerian flow with significant advection are obtained.

\begin{figure}
\epsfig{figure=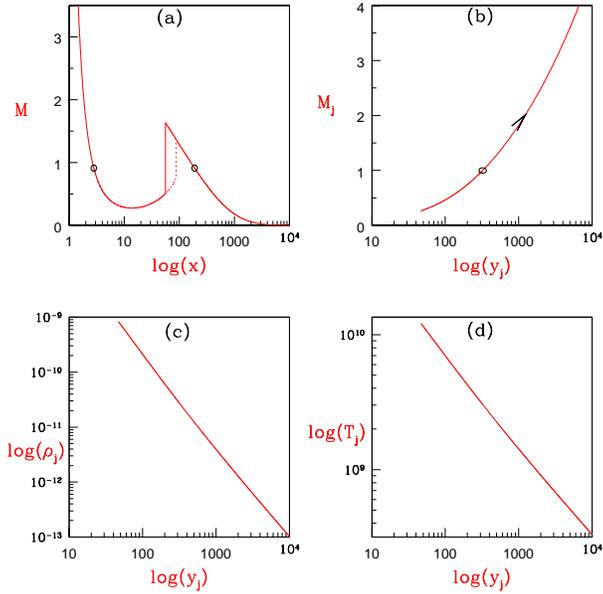,height=8.cm,width=8.0cm,angle=0} 
\caption{Mach number distribution for self-consistent accretion-ejection solution (a \& b).
The disc solution (a), and the jet (b) are plotted for accretion flow parameters $E=0.001$, $\lambda_0=1.542$ and $\alpha=0.05$.The shock
in accretion is shown by the vertical jump at
$x_s=56.236$ (a).
The density distribution $\rho_j$ (c) and the temperature $T_j$ distribution of the jets are also
plotted (d). 
The sonic points are marked by open circles (a \& b). The dotted vertical line shows the position
of the shock if $R_{\dot m}=0$. 
The jet flow parameters ${\cal E}_j={\cal E}_+=1.084\times 10^{-3}$, $\lambda_j=\lambda_+=1.699$.}
\label{lab:fig7}
\end{figure}

The effect of $\alpha$ on shock location $x_s$ is an interesting issue. Numerical simulations show that,
for fixed outer boundary condition,
$x_s$ expands to larger distances with the increase of $\alpha$ \citep{lmc98,ldc11}, while analytically
\citet{cd07} showed that for the same outer boundary condition, $x_s$ shifts closer to the horizon
with the increase in $\alpha$.
Although, the viscosity prescription of \citet{cd07} and the simulations are different, namely
the former chose the stress to be proportional to total pressure, while in simulations the stress is proportional
to the shear, still viscosity reduces angular momentum, and we know for lower angular momentum
if the shock forms, it should form closer to the black hole! Since the viscosity prescription in this paper is
similar to the simulations, we should be able to answer the dichotomy.
So a concrete question may arise, if viscosity
is increased, does the $x_s$ expands to larger distances or, contracts to a position closer to the horizon?
If the viscosity acts in a way such that the $\lambda_+$ (immediate post-shock $\lambda$) is less than its inviscid
value at the shock then $x_s$ will move closer to the horizon. However, such simple minded reasoning may fail,
if shocks exists, then the post shock flow being hotter would transport more efficiently
than the immediate pre-shock flow. So although, $\lambda_-<\lambda(x_{inj})$, it is not necessary
$\lambda_+$ will be less than $\lambda(x_{inj})$.  
We scourged the parameter space to find the answer, and in the following we present the explanation.

\begin{figure}
\epsfig{figure=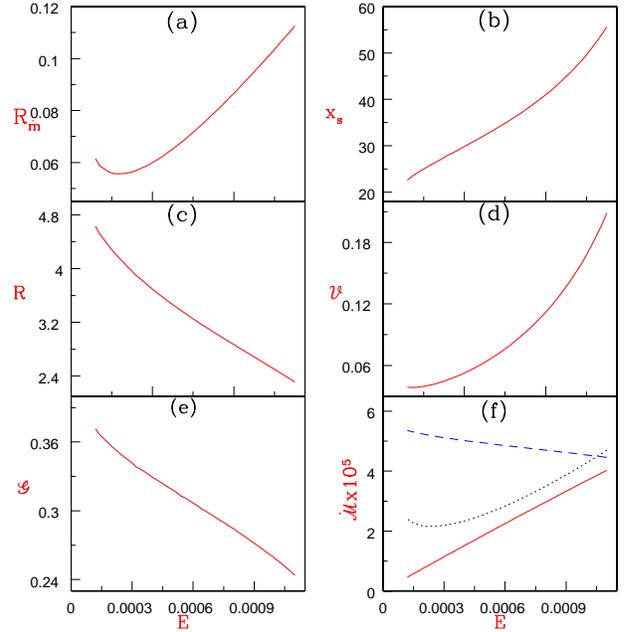,height=8.5cm,width=8.5cm,angle=0} 
\caption{Various shock and jet
quantities like $R_{\dot m}$ (a), $x_s$ (b), $R$ (c), ${\cal V}$ (d),
${\cal G}$ (e), and ${\dot {\cal M}}$ (f) are plotted with $E$
for $\alpha=0.05$ and $\lambda_0=1.54$.}
\label{lab:inouflo_fig}
\end{figure}

Let $x_b$ be the distance at which the $\lambda$ distribution of the viscous solution is coincident
with the $\lambda$ value of the inviscid solution, let us further assign $\lambda_b=\lambda(x_b)$.
It is to be remembered that $\lambda_0=\lambda_b$ for the inviscid ($\alpha=0$) solution,
but $\lambda_0<\lambda_b$ for viscous solution because viscosity reduces the angular momentum. In all
the simulations done on viscous flow in the advective regime
referred in this paper, one starts with an inviscid solution (since analytical solutions are readily
available), and then the viscosity is turned on, keeping the values at the outer boundary fixed.
In other words, it is this $x_b$ that is called the outer boundary in numerical simulations,
and generally, $x_b \lsim \rm{few}\times100$ since it is computationally expensive to simulate a large domain from just outside the horizon to
few$\times 1000~r_g$ and still retain required resolution to achieve intricate structures
in the accretion disc.
In Figs. \ref{lab:adaf_fig}a-c, we compare the $\lambda(x)$ of shocked accretion flows starting
with the same $E=10^{-4},~\&~\lambda_b=1.68$ for various viscosity parameters
such as, $\alpha=0$ (solid), $\alpha=0.0075$ (dotted), $\alpha=0.015$, but for different points of coincidence,
\eg $x_b= 2220$ (5a), $x_b=3660$ (5b), and $x_b=5230$ (5c).
The vertical solid line is the location of the shock for the inviscid flow. Although, we match the $\lambda_b$
at $x_b$ of the viscous and inviscid solutions, we still integrate outwards upto $x_{inj}=10^4$. Since $x_{inj}>>1$, this distance may be considered as
the size of the disc. In Figs. \ref{lab:adaf_fig}a-c, we show that depending upon the choice of $x_b$, the outer boundary conditions can be remarkably different. If $x_b$ is short (Fig. \ref{lab:adaf_fig}a),
then flow with higher $\alpha$ will be able to match $\lambda_b$ at the same $x_b$, only if
one starts with much higher $\lambda_{inj}$, in other words the gradients will be stiffer. Consequently,
the increase of
$\alpha$ will create higher
$\lambda_+$, which will result in the increase of $x_s$. In case
$x_b$ is large (Fig. \ref{lab:adaf_fig}c), the gradients are smoother and the resulting
$\lambda_{inj}$ will be approximately similar for any value of $\alpha$. Hence as $\alpha$ is increased,
$\lambda_+$ would decrease and consequently $x_s$ will decrease too. In Fig. \ref{lab:adaf_fig}a, the parameters are
$\lambda_{inj}=\lambda_b=1.68$ for
inviscid or $\alpha=0$ (solid), $\lambda_{inj}=1.769$ for $\alpha=0.0075$ (dotted) and
$\lambda_{inj}=1.858$ for $\alpha=0.015$ (dashed). Since $x_b=2220$ is short, the gradients
are steeper, and as discussed above, in such cases $x_s$ increases with increasing $\alpha$.
While in Fig. \ref{lab:adaf_fig}c, $\lambda_{inj}=1.744$ for $\alpha=0.0075$ (dotted) and $\lambda_{inj}=1.807$
for $\alpha=0.015$ (dashed), where $x_b=5230$ is larger, the shock $x_s$ decreases with increasing $\alpha$.
Since for shorter values of $x_b$, $x_s$ increases with the increase of $\alpha$, and for longer
$x_b$, $x_s$ decreases with the increase of $\alpha$, so a limiting $x_b$ should exist
for which $x_s$ will neither increase or decrease with the increase of $\alpha$. In Fig. \ref{lab:adaf_fig}b, we show that for $x_b=3660$, the shock neither increase or decrease with
the increase of $\alpha$.

In Figs. \ref{lab:adaf_fig}d \& e, we plot the limiting $x_b$ as a function of $E$ for
$\lambda_0=1.68$
(d) and $\lambda_0=1.65$ (e). The domain name `1' corresponds to any $x_b$ at which the $x_s$ will decrease
and `2' signifies the domain where at any $x_b$, $x_s$ increases with $\alpha$.
Hence Fig. \ref{lab:adaf_fig}a lies in domain 2, Fig. \ref{lab:adaf_fig}b on the curve,
and Fig. \ref{lab:adaf_fig}c on domain 1 of Fig. \ref{lab:adaf_fig}d.
Incidentally,
if the outer boundary condition for all the advective solutions start with the Keplerian angular momentum,
then the shock location
$x_s$ decreases with the increase of $\alpha$. Since the numerical simulations are usually performed 
with a shorter radial extent ($x_b\sim \rm{few}\times 10$--- few$\times 100$),
this is similar to the case $x_b$ lying in the domain 2. As a result, earlier simulations of
advective accretion flows have reported
the increase of $x_s$ with the increase of $\alpha$. Hence we may conclude that the increment of $x_s$
with the increase of $\alpha$ is an artifact of faulty assignment of outer boundary condition in the simulations.

In Fig. \ref{lab:sokparam_fig}, the $E$---$\lambda_0$ parameter space for standing
accretion shock has been plotted for various viscosity parameters, $\alpha=0$, $0.05$, $0.1$,
$0.15$, $0.2$,
$0.25$, $0.3$. Since viscosity will in general reduce the angular momentum along the flow,
$\lambda_0$ should decrease with the increase of $\alpha$. As a result,
the shock parameter space shift to the lower end of the $\lambda_0$ scale.
One may compare the RH shock parameter space with that of the
isothermal shock space \citep{dbl09}. For all possible boundary conditions, RH shocks
may be obtained upto $\alpha=0.3$. For values outside the bounded regions there are no standing shocks,
although transient and oscillating shocks may exist. It is to be remembered though, the parameter
space shown here corresponds to inner boundary. In the outer boundary $\lambda$ might be much higher
for viscous flow as has been shown in Figs. \ref{lab:adsh_fig}a-e. 

\begin{figure}
\epsfig{figure=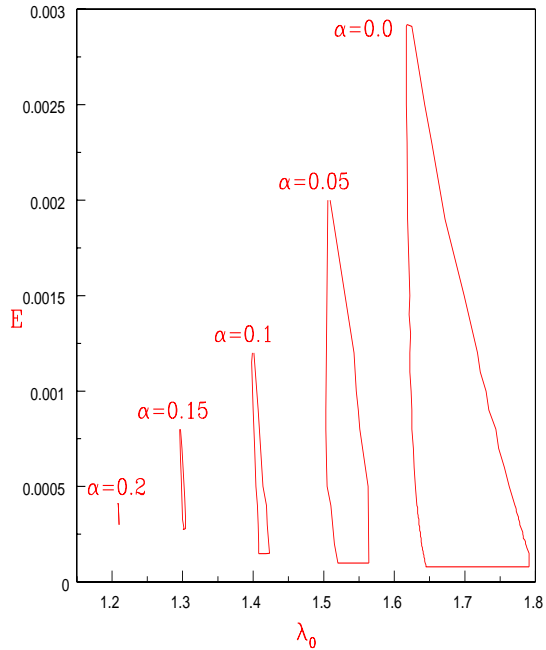,height=10.cm,width=8.cm,angle=0} 
\caption{$E-\lambda_0$ parameter space of the standing adiabatic shock with Shakura-Sunyaev viscosity parameter
$\alpha=0.0$, $0.05$, $0.1$, $0.15$, $0.2$ marked on the figure, and in presence of massloss.}
\label{lab:fig9}
\end{figure}

\subsection{Inflow-Outflow solutions}

In the previous sub-section we discussed all possible solutions that one can have
in presence of viscosity prescription given by Eq. (\ref{strs.eq}), and non-dissipative or RH shocks.
In this section we present the self consistent inflow-outflow solutions. In presence of massloss,
the mass conservation equation across the shock will be modified in the form of Eq. (\ref{mfs.eq}).
All the steps
mentioned in section 3.2 are followed to compute the self-consistent inflow-outflow solution.
In Fig. \ref{lab:fig7}a-d, we present one case of global accretion-ejection
solution. In Fig. \ref{lab:fig7}a, the accretion solution in terms
of $M(x)$ or the Mach number is presented, in Fig. \ref{lab:fig7}b, the jet Mach number $M_j$ 
is plotted with the height $y_j$ of the jet from the equatorial plane of the disc, where the radial coordinate of the
jet is $r_j=\sqrt{x^2_j+y^2_j}$. The sonic points are marked with open circles.
The inner boundary for the accretion solution is $E=0.001$, $\lambda_0=1.542$. The specific energy and the
angular momentum at the shock are ${\cal E}_+=1.084\times 10^{-3}$, $\lambda_+=1.699$.
The collimation parameter of the jet
(see, Chattopadhyay 2005) at its sonic point is $x_{jc}/y_{jc}\sim 0.24$,
and hence the spread is quite moderate. The dotted vertical line is the position of the
shock when $R_{\dot m}=0$, and the solid vertical line is the position of the
shock after the mass-outflow rate has been computed. Clearly, since excess thermal gradient force
in the post-shock disc drives bipolar outflows, it reduces the pressure, and hence
to maintain the total pressure balance across the shock, the shock front moves closer
to the horizon. The relative mass outflow rate computed for the particular case depicted in
Figs. \ref{lab:fig7}a-d, is  
$R_{\dot m}=0.1044$.  In Fig. \ref{lab:fig7}c, we plot the density $\rho_j$ of the jet
derived for ${\dot M}=0.1M_{\rm Edd}$ for a black hole of $M=10M_{\odot}$. In Fig. \ref{lab:fig7}d,
we plot the temperature $T_j$ of the jet. As is expected the jet is hot near its base but falls to
low temperatures at larger distances away, while the density also falls as is expected of
a transonic jet.

\begin{figure}
\epsfig{figure=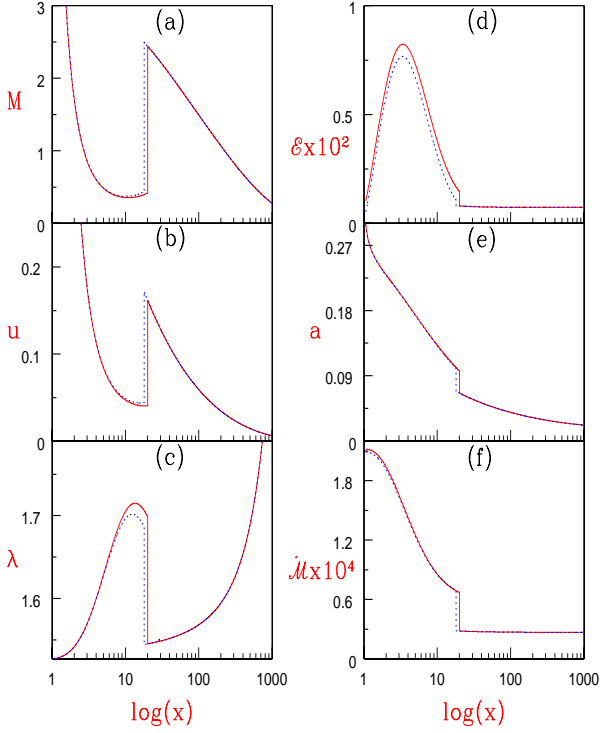,height=10.cm,width=8.cm,angle=0} 
\caption{Comparison of accretion solutions with non-dissipative
shock \ie $f_e=0$ (solid),
and dissipative shock $f_e=0.1$ (dotted). The inner boundary condition for $f_e=0$
are $E,~\lambda_0=7.373\times 10^{-4},~1.527$ (solid), and for $f_e=0.1$ are $E_+,~\lambda_0=2.5\times 10^{-4},
~ 1.527$ (dotted). In case of the dotted curve $E_-=7.373\times 10^{-4}$. For $f_e=0$, $x_s=20.15$ , \&
$R_{\dot m}=0.1133$ (solid),
and for $f_e=0.1$, $x_s=18.04$ \& $R_{\dot m}=0.0755$ (dashed). Various flow variables are
$M$ (a), $u$ (b), $\lambda$ (c), ${\cal E}$ (d), $a$ (e), and ${\dot {\cal M}}$ (f).}
\label{lab:fig10}
\end{figure}

Figures \ref{lab:inouflo_fig}a-f, are plotted for the fixed values of $\alpha=0.05$
and $\lambda_0=1.54$. Various quantities $R_{\dot m}$ (a), $x_s$ (b), $R$ (c), ${\cal V}$ (d),
${\cal G}$ (e), and ${\dot {\cal M}}$ (f) are plotted with $E$, which means, we are studying how
$R_{\dot m}$ change with the change of inner boundary condition.
Generally the $R_{\dot m}$ is few percent of the injected accretion rate but it can go upto more than $10\%$ of ${\dot M}_-$ for high $E$.
Although in this plot, our main interest is to quantify the relative mass outflow rate $R_{\dot m}$ with $E$
for given value of $\alpha$ and $\lambda_0$, we have plotted all the essential quantities
that would contribute in driving a part of the post-shock matter as outflows.
The shock location $x_s$ increases with the increase of $E$, and for increasing
$x_s$, $R$ should decrease. 
Since $R_{\dot m}$ is combination of $R$, ${\cal V}$ and ${\cal G}$,
even for falling $R$ and ${\cal G}$, the mass outflow rate increases since it is being
compensated by the increase of ${\cal V}$. Interestingly,
none of the quantities show a monotonic variation. It is important to note, all the three
parameters $R$, ${\cal V}$ and ${\cal G}$ represent the jet to disc connection and not the actual driving.
The real drivers are however, the post-shock specific energy ${\cal E}_+$, and the jet base cross-section ${\cal A}_s(\equiv {\cal A}(x_s))$. 
Higher ${\cal E}_+$ means hotter flow at the jet base, and therefore the thermal driving will be more. This is complemented by the cross-section ${\cal A}_s$ of the jet base. Higher ${\cal E}_+$ would drive more matter into the jet
channel but will be limited by the cross-sectional area. 
In Fig. \ref{lab:inouflo_fig}f, we plot the pre-shock entropy-accretion rate ${\dot {\cal M}}_-$
(solid), the post-shock entropy-accretion rate ${\dot {\cal M}}_+$ (dashed) and the 
jet entropy-accretion rate ${\dot {\cal M}}_j$ (dotted), as a function of $E$, and it is quite evident
that the pre-shock entropy is less than both ${\dot {\cal M}}_j$ and ${\dot {\cal M}}_+$.
Moreover, when the value of ${\dot {\cal M}}_j$ is high, $R_{\dot m}$ is found to be high too,
which vindicates that matter would prefer to flow through channels with higher entropy.

From Eq. (\ref{mfs.eq}), we know that ${\dot M}_+<{\dot M}_-$ if $R_{\dot m}\neq0$. Therefore,
the post-shock pressure would be reduced. This would cause $x_s$ to decrease as shown
in Fig. \ref{lab:fig7}a [also see \citet{cd07}].
Furthermore, the massloss from the post-shock flow and consequent reduction in pressure would also modify shock parameter space. In Fig. \ref{lab:fig9}, the bounded regions
in the $E-\lambda_0$ space, represents the shock parameter space for $\alpha=0$, $0.05$,
$0.1$, $0.15$ and $0.2$, marked on the figure. Compared to the shock parameter space in absence of massloss (Fig. \ref{lab:sokparam_fig}), the parameter
space in presence of massloss gets reduced and beyond $\alpha=0.2$ the standing shock seems to vanish. 
Moreover, the shock parameter space shows that the standing shock do not seem to exist
for very low $E$. Non-existence of standing shock ofcourse do not imply non-existence of non-steady or
oscillatory shocks. 

\begin{figure}
\epsfig{figure=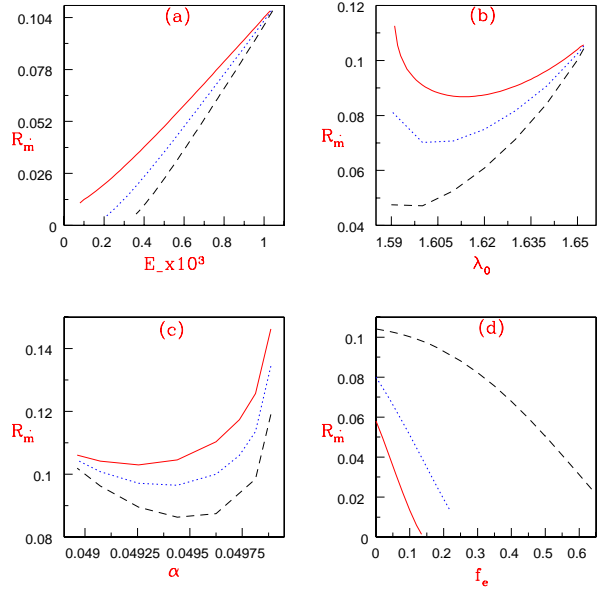,height=8.cm,width=8.cm,angle=0}  
\caption{Relative mass outflow rate $R_{\dot m}$ as a function of (a) $E_-$ for fixed values of $\alpha=0.02$,
$\lambda_0=1.65$; (b) $\lambda_0$ for fixed values of $E_-=0.001$ and $\alpha=0.02$; (c) $\alpha$
for fixed values of $E_-=0.001$ and $\lambda_{inj}=70.6$
at $x_{inj}=10^4$; where all the plots are for $f_e=0$ (solid), $f_e=0.05$ (dotted) and $f_e=0.1$ (dashed).
(d) $R_{\dot m}$ as a function of $f_e$ for fixed values of $E_-=0.001$ and $\lambda_0=1.65$
and each curves are for $\alpha=0$ (solid), $\alpha=0.01$ (dotted) and $\alpha=0.02$ (dashed).}
\label{lab:fig11}
\end{figure}

\subsection{Massloss from the dissipative shocks}

\citet{ct95} and later \citet{mc10} had shown that the post shock hot flows can produce the high
energy photons easily
by inverse-Comptonizing the soft photons and reproduced the observed spectra from variety of objects.
If indeed the observed spectra and luminosity
can be reproduced from the post-shock flow, then the grand energy will not
be conserved across the shock and will be given by Eq. (\ref{ef.eq}), and the dissipated energy
can be radiated away. In Fig. \ref{lab:fig10}a-f, various flow variables $M$ (a), $u$ (b), $\lambda$ (c), ${\cal E}$ (d), $a$ (e), and ${\dot {\cal M}}$ (f) are plotted for $f_e=0$ (solid) and $f_e=0.1$
(dotted). For $f_e=0$ (solid), the solutions are plotted for flow parameters
$E,~\lambda_0=7.373\times 10^{-4},~1.527$. For $f_e=0$, we have $E=E_+=E_-$.
However, for $f_e>0$, $E_+<E_-$. So for $f_e=0.1$ (dotted), the inner boundary is
represented by $E_+,~\lambda_0=2.5\times 10^{-4},
~ 1.527$, while the pre-shock $E_-=7.373\times 10^{-4}$.
The post-shock specific energy or Bernoulli parameter, temperature, angular momentum, and entropy
of the solution with dissipative shock (dotted) are lower than those corresponding to the
non-dissipative shock (solid).
Since a part of the thermal energy gained through shock is spent in powering jets and to produce
radiation, the shock front moves closer to the black hole.
The relative mass outflow rate or $R_{\dot m}$ is lower for dissipative shocks, because part of the thermal energy
of the post-shock disc
is lost through radiation. However, the thermal energy lost as radiation can still contribute to jet power
if those photons deposit momentum onto the jets \citep{cc02,cdc04,c05}. 
\begin{figure}
\epsfig{figure=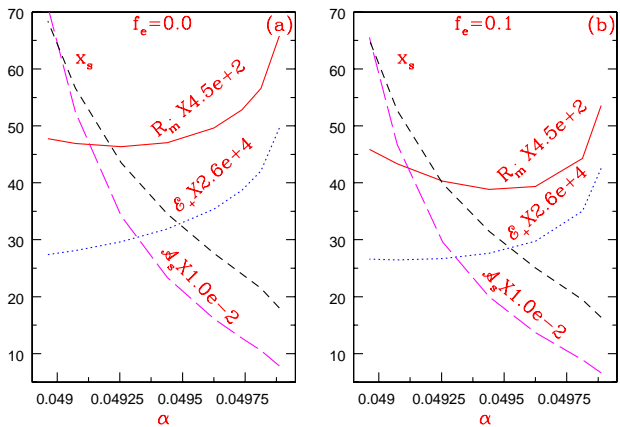,height=8.5cm,width=8.5cm,angle=0}  
\caption{Relative mass outflow rate $R_{\dot m}$ (solid), $x_s$ (dashed), ${\cal E}_+$ (dotted),
and ${\cal A}_s$ (long-dashed) as a function of $\alpha$ for $f_e=0$ (a) and
$f_e=0.1$ (b) for fixed values of $E_-=0.001$ and $\lambda_{inj}=\lambda_K(x_{inj})=70.6$
at $x_{inj}=10^4$.}
\label{lab:fig12}
\end{figure}

If $f_e\neq 0$ and shocks are present, then the accretion solutions are obtained either for $E_+,~\lambda_0,~\alpha,~f_e$,
or equivalently,
$E_-,~\lambda_0,~\alpha,~f_e$, or, $E_-,~\lambda_{inj},~\alpha,~f_e$. If $f_e=0$ then
$E=E_+=E_-$, therefore any of the following two sets would suffice
$E,~\lambda_0,~\alpha$, or, $E,~\lambda_{inj},~\alpha$. 
Since the outflows are launched from the post-shock
flow, the mass outflow rates are computed only from the shocked accretion flows,
and we are interested to find the dependence of $R_{\dot m}$ on each of the above mentioned
parameters.
In Fig. \ref{lab:fig11}a,
we plot $R_{\dot m}$ as a function of $E_-$ when the other parameters are fixed
at $\alpha=0.02$, $\lambda_0=1.65$ where each of the curve are for
$f_e=0$ (solid), $f_e=0.05$ (dotted) and $f_e=0.1$ (dashed).
With the increase of $E_-$, the post-shock specific energy ${\cal E}_+$ increases which drives
more matter as outflow, and hence $R_{\dot m}$ increases. 
However, for any given $E_-$, $R_{\dot m}$ decrease with the increase of $f_e$, \ie
the mass outflow decreases with the increase of the thermal energy dissipation. 
Although for very high $E_-$, the separation of the curves decreases, this shows that
flows starting with higher energy would have enough thermal energy to drive significant
outflows even for $f_e\neq0$. Now, let us fix $E_-$ but vary $\lambda_0$ in Fig. \ref{lab:fig11}b.
$R_{\dot m}$ is plotted with $\lambda_0$ for $f_e=0$ (solid), $f_e=0.05$ (dotted) and $f_e=0.1$ (dashed).
The fixed parameters are $E_-=0.001$, and $\alpha=0.02$, and clearly $R_{\dot m}$ is not a monotonic
function of $\lambda_0$.
Increasing $\lambda_0$ would increase $\lambda_+$ and therefore increase
$x_s$ but would decrease ${\cal E}_+$. The increase of $x_s$, increases the jet base cross-section.
The competition between ${\cal E}_+$ and ${\cal A}$ causes a dip in $R_{\dot m}$ for moderate values of $\lambda_0$.

However, $R_{\dot m}$ decreases with the increase of $f_e$. %
In Fig. \ref{lab:fig11}c, we fix the outer boundary condition, namely, $E_-=0.001$, $\lambda_{inj}=70.6$
at $x_{inj}=10^4$, and $R_{\dot m}$ is plotted
as a function of $\alpha$, where each of the curves are for $f_e=0$ (solid), $f_e=0.05$ (dotted) and $f_e=0.1$ (dashed).
Since for flows starting with such high $\lambda_{inj}$
there would be no accretion shock solution for low $\alpha$, therefore one observes outflows
only beyond a critical value of $\alpha$. We know (see Fig. \ref{lab:adaf_fig}c) that for flow starting with
the same outer boundary condition, the shock location decreases with the increasing $\alpha$.
Decrease in $x_s$ would mean decrease in ${\cal A}_s$. Since $x_s$ decreases, it means
the post-shock energy increases due to viscous dissipation \ie ${\cal E}_+$ increases.
Increase in ${\cal E}_+$ would drive more matter into the outflow channel. Therefore, the decrease in
${\cal A}_s$ is dominated by the increase in ${\cal E}_+$, and
so $R_{\dot m}$ increases with $\alpha$ for a fixed outer boundary condition.

However, $R_{\dot m}$ decreases with the increase in $f_e$, which also shows that these jets are thermally driven.
In Fig. \ref{lab:fig11}d, $R_{\dot m}$ is plotted with $f_e$ for
$\alpha=0$ (solid), $\alpha=0.01$ (dotted) and $\alpha=0.02$ (dashed). The fixed parameters
are $E_-=0.001$ and $\lambda_0=1.65$. It was also found out that there are critical $f_e$ beyond
which no
standing shock conditions are satisfied, and they are $f_{ec}=0.135$ for $\alpha=0$ (solid), $f_{ec}=0.22$ for $\alpha=0.05$
(dotted) and $f_{ec}=0.645$ for $\alpha=0.1$ (dashed). It is obvious that $f_{ec}$ increases 
with the increase of $\alpha$.

 It is to be noted that $R_{\dot m}$ is the fraction of matter which is shock heated and
ejected out as bipolar outflows. 
To properly understand the role of shock in driving bipolar outflow, we consider
accretion solutions with the same outer boundary condition, or the case presented in Fig. \ref{lab:fig11}c. In Fig. \ref{lab:fig12}a, we plot various quantities across the shock \eg, $R_{\dot m}$ (solid), ${\cal E}_+$ (dotted),
$x_s$ (dashed), and ${\cal A}_s$ (long dashed), as a function of $\alpha$, and for $f_e=0$, \ie corresponding to the solid plot
of Fig. \ref{lab:fig11}c. In Fig. \ref{lab:fig12}b, we plot $R_{\dot m}$ (solid), ${\cal E}_+$ (dotted),
$x_s$ (dashed), and ${\cal A}_s$ (long dashed), as a function of $\alpha$, and for $f_e=0.1$, \ie corresponding to the dashed plot
of Fig. \ref{lab:fig11}c. Increasing $\alpha$ in accretion with the same outer boundary condition, will decrease $\lambda_+$ and therefore
decrease $x_s$ (dashed). Decrease in $x_s$ implies that the shock is formed closer to the black hole,
where the viscous dissipation would be more \ie ${\cal E}_+$ will be higher (dotted).
The jet velocity is small at the jet base, 
so large ${\cal E}_+$ means hotter flow and stronger driving of the outflow \ie higher $R_{\dot m}$. %
Figure \ref{lab:fig12}b
shows that, as the shock dissipation $f_e$ is increased, at certain $\alpha$, ${\cal E}_+$ is reduced while the decrease in ${\cal A}_s$ is marginal, this causes $R_{\dot m}$ initially to decrease with $\alpha$,
but eventually starts to increase as ${\cal E}$ increases appreciably. %

\begin{figure}
\epsfig{figure=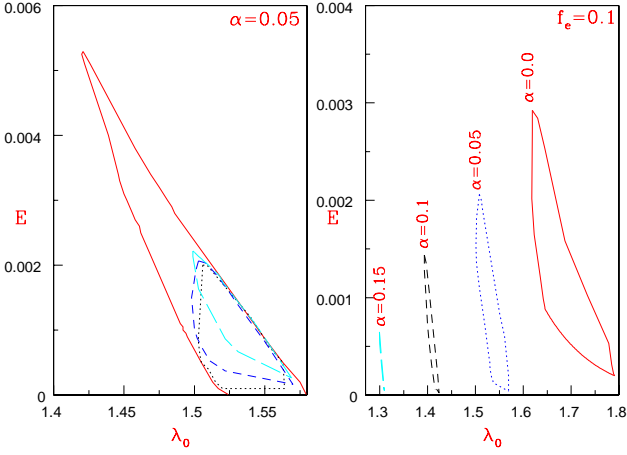,height=8.5cm,width=8.5cm,angle=0} 
\caption{$E-\lambda_0$ shock parameter space for
(a) $\alpha=0.05$, and each curve is for $f=1,~f_e=0$ (solid),
$f>1,~f_e=0$ (dotted), $f>1,~f_e=0.05$ (dashed), and $f>1,~f_e=0.2$ (long dashed),
and (b) $f>1,~f_e=0.1$, and each curve is for $\alpha=0.0$ (solid),
$0.05$ (dotted), $0.1$ (dashed), and $0.15$ (long dashed).}
\label{lab:fig13}
\end{figure}

In Fig. \ref{lab:fig13}a, we plot the $E-\lambda_0$ shock parameter space for flows
with $f=1,~f_e=0$ (solid),
$f>1,~f_e=0$ (dotted), $f>1,~f_e=0.05$ (dashed), and $f>1~f_e=0.2$ (long dashed).
The viscosity parameter for this figure is $\alpha=0.05$.
It is to be remembered $f=1$ means no massloss and $f>1$ means presence of bipolar outflows.
Therefore $f=1,~f_e=0$ (solid), implies the shock parameter space for RH shocks with no massloss
and no dissipation at the shock, while $f>1,~f_e=0$ (dotted) implies non dissipative shock
but massloss is present. Similarly, $f>1,~f_e=0.05$ (dashed) and $f>1,~f_e=0.2$ (long dashed)
shock parameter space for dissipative shocks of various strength and in presence of massloss.
The parameter space for standing shock shrinks with the increase of $f_e$.
In Fig. \ref{lab:fig13}b, we plot the shock parameter space for fixed $f_e=0.1$ and various
values of $\alpha=0$ (solid), $\alpha=0.05$ (dotted), $\alpha=0.1$ (dashed) and $\alpha=0.15$
(long dashed). We know from Fig. \ref{lab:sokparam_fig}, that with the increase of $\alpha$
the shock parameter space shrinks, however, in presence of massloss, flows with low $E$
seems to show no standing shock. In Fig. \ref{lab:fig13}b, we see that although the parameter space for shock
shrinks but low $E$ flow again exhibit standing shock, which signifies that
the mass outflow rate decreases with the increase of $f_e$.

\section{Discussion and conclusion}

Quasars and micro-quasars may show strong jets, and these outflows are correlated with the
spectral state of the object. Therefore, quantitative estimate of the generation of these outflows
are required, especially, the relation between the mass outflow rates and the viscosity parameter
needs to be ascertained, this is because the viscosity
will determine the disc spectral states. Most of these estimates are available for inviscid flow, or special viscosity
prescription or shock condition, but not for the most general shock condition, \ie partially dissipative
shock. 

In this paper, our main concern has been to estimate the thermally driven bipolar outflows
from shocked accretion discs around black holes. Since accretion disc should
posses some amount of viscosity, a proper understanding of viscous accretion disc
is required. As outflows are generated from the post-shock disc
and high energy photons should also emerge from the post-shock disc,
an investigation of estimating the outflow rate and correlating it with the dissipation 
parameters is a must requirement.
In order to estimate the outflows correctly,
a proper understanding of the accretion process has to be undertaken, and we presented all possible
accretion solutions without massloss in section 4.1,
including the shocked and shock free accretion solutions and the dependence of these solutions on viscosity
parameter. While doing so we have compared solutions with same inner and outer
boundary conditions, and have shown that these differences produces a significant difference in
interpreting the results.

For solutions with the same
inner boundary conditions (\ie Figs. \ref{lab:eov_fig}a-p),
we found two critical viscosity parameters, the first one being the onset of shock ($\alpha_{\rm cl}$),
and the other being
the one above which standing shock disappears ($\alpha_{\rm cu}$) and
generates a shock free global solution which is wholly subsonic except close to the horizon, or the ADAF
type solution. Simultaneous existence of
both $\alpha_{\rm cl}$ and $\alpha_{\rm cu}$ will depend on whether $\lambda_0$ is small enough
to produce a corresponding O-type inviscid solution.
If inviscid solution is already shocked then only $\alpha_{\rm cu}$ will exist, while
if the inviscid solution is I or W type then neither $\alpha_{\rm cl}$ nor $\alpha_{\rm cu}$ exists.
For solutions with the same outer boundary condition, \ie flow starting with
same $E$, $\lambda_{inj}=\lambda_K(x_{inj})$ at some injection radius $x_{inj}$, \eg Fig. \ref{lab:adsh_fig}a-e, there would be an additional
critical viscosity parameter $\alpha_1$, which would allow a global solution connecting the horizon
and the outer boundary $x_{inj}$.
We have also confirmed that fluids in such cases, $x_s$ would decrease with the increase of $\alpha$.
The decrease of $x_s$ with the increase of $\alpha$
is interesting.
\citet{ct95} for the first time argued that the post-shock disc is the elusive Compton cloud, which
inverse-Comptonizes the pre-shock soft photons to produce power law tail. If the shock remains strong
then we have the canonical hard state and when the shock becomes weak or disappear we have the
canonical soft state. Moreover, \citet{msc96} showed that if the shock oscillates, it does with a frequency
$\omega \sim x^{- \beta}_s$, where $\beta =1~\rightarrow 3/2$. If the shock oscillates then the hard radiation
from it would oscillate with the same frequency and could explain the mHz to
few tens of Hz QPO observed in stellar mass black hole candidates. Outburst phase in microquasars starts with low frequency QPOs in hard state, but as the
source moves to intermediate states the QPO frequency increases to a maximum and then goes down in the declining phase, a fact well explained by approaching oscillating $x_s$ with the increase of viscosity \citep{cdp09}.
Our steady state model also shows that if every other conditions at the outer boundary is same, then $x_s$
decreases with the increase of $\alpha$, so we expect with the increase of viscosity the shock oscillation
too will increase, and therefore the shock oscillation model of QPO seems to follow observations.

We have computed the mass outflow rate from both non-dissipative as well as dissipative shocks.
The mass outflow rate is always of higher entropy than the pre-shock disc, which shows that
mass-outflow rate is a natural consequence of a shocked accretion disc, a fact readily supported
by various multi-dimensional simulations. In this connection we would like to comment that
unlike \citet{cd07,dc08}, we have corrected the jet cross sectional area with the projection effect
onto the jet streamline.  
We showed that $R_{\dot m}$ generally
decreases with $f_e$. We also showed that the parameter space is significantly modified due to the
presence of massloss and dissipation at the shock. When dissipative shocks are included we do see that
the relative mass outflow rate decreases which are also to be expected. Increasing dissipation would make
the shocks weaker, which can be identified with the softer spectral state, and the accretion disc in the soft
state will give weaker or not outflow \citep{gfp03}. 

Comparison of steady shock parameter space Figs. (\ref{lab:sokparam_fig},\ref{lab:fig9}), suggests that massloss
may trigger an instability. It shows that parts of parameter space which produced steady shocks in absence of
of massloss, do not show steady shocks in presence of massloss. This is because
the post-shock pressure gets reduced as it looses mass, and the shock moves closer to the black hole
(Fig. \ref{lab:fig7}) to regain the momentum balance. However, this is not always possible, and the shock may
oscillate or get disrupted altogether, therefore, there should be a massloss induced instability as well.
We have also shown that the main driver for the bipolar outflow is the energy gained through shock. Interestingly,
the mass outflow rate generally increase with the increasing viscosity parameter. Since the
the shock location also decreases with the increasing $\alpha$ for identical outer boundary condition,
there is a possibility that the QPOs and mass outflow rates be correlated --- an issue we will pursue
in fully time dependent studies.

It is interesting to note that the dissipation parameter used in this work has been assumed constant, however
this would actually depend on accretion rates and the size of the post-shock region. As the accretion rate changes
the total photons generated would change, and similarly as $x_s$ changes the fraction of photons
intercepted by post-shock disc as well as its optical depth would change, this would render $f_e$ variable.

It has been observationally established that steady jets are observed from black hole candidates, when the spectrum of the disc is in the hard state \citep{gfp03}. In our model, the presence of strong shock is
similar to the hard state. We have shown that with the increasing $\alpha$, we get stronger jet, however evolution of jet states with spectral states 
is a time-dependent phenomenon.  Moreover, the phenomena of QPO and growth or decay of QPO are time dependent phenomenon too.
Our conjecture that the QPO will be correlated with the jet states can only be vindicated through a fully
time dependent study, which is beyond the scope of this paper.
Furthermore, since we concentrated on the effect of viscosity on accretion disc and
outflows, therefore, magnetic field
and other realistic cooling processes have been ignored. If cooling processes are considered then a direct comparison with the
observation will be possible. As has been noted that a little bit of magnetic field will have an important effect on the
dynamics of the flow \citep{p05}, however transonicity criteria will be important too. Because a magnetized flow 
may posses fast, slow or Alfvenic waves, and the number of sonic points may increase \citep{tgfrt06}.
If magnetized flows admit shocks, then the shock produced, may be even more robust. And
this hydrodynamic model might well act as the simpler version of the magneto fluid model.
We are working on dynamics of magnetized flows and would be reported elsewhere.

The concrete conclusions we draw from this paper is the following.
Viscosity is important and affects the accretion solutions both quantitatively and qualitatively.
Shock in accretion can be obtained for fairly high viscosity parameter. Shocks naturally produces
outflows, and for fixed outer boundary conditions of the disc, shock location decreases but
mass outflow rate generally increases. This augurs well for the model as this is exactly
observed in hard to intermediate hard spectral transitions. However, in presence of dissipative
shocks the mass outflow rate decreases. Over all we see that $R_{\dot m}$ may vary between few \% to more than 10 \%, although in presence of
dissipative shocks, the estimate of $R_{\dot m}$ is about few \%. We have also computed the shock parameter space
for accreting flows without mass loss and dissipation, with massloss but no dissipation and with massloss
and dissipation and have shown that the parameter space for steady shocks shrinks with increasing dissipation.

%\section*{Acknowledgments}

 \label{lastpage}
%% %%\bibliographystyle{alpha} %% give your .bst file
%% %%\bibliography{ms_joshi}   %% give your .bib file

\end{document}